\documentclass[sigconf, nonacm]{acmart}
\pdfoutput=1

\newcommand\vldbdoi{XX.XX/XXX.XX}
\newcommand\vldbpages{XXX-XXX}
\newcommand\vldbvolume{14}
\newcommand\vldbissue{1}
\newcommand\vldbyear{2020}
\newcommand\vldbauthors{\authors}
\newcommand\vldbtitle{\shorttitle} 
\newcommand\vldbavailabilityurl{http://vldb.org/pvldb/format_vol14.html}
\newcommand\vldbpagestyle{plain}

\usepackage{amsthm}
\usepackage{times}
\usepackage{graphicx}
\usepackage{balance}  
\usepackage{cite}
\usepackage{amsmath,amsfonts}
\usepackage{algorithm}
\usepackage[noend]{algorithmic}
\usepackage{textcomp}
\usepackage{xcolor}
\usepackage{setspace}
\usepackage{multirow}
\usepackage{multirow}
\usepackage{url}
\usepackage{epsfig,subfigure}
\usepackage{xspace}

\usepackage{graphicx}
\usepackage{balance}  
\usepackage{booktabs}
\usepackage{xspace}
\usepackage{epsfig,subfigure}
\usepackage{epstopdf}
\usepackage{upgreek}

\usepackage{url}
\usepackage{multirow}

\newtheorem{problem}{Problem}
\newtheorem{definition}{Definition}
\newtheorem{observation}{Observation}
\newtheorem{example}{Example}

\newcommand{\squishlisttight}{
\begin{list}{$\bullet$}
	{ \setlength{\itemsep}{0pt}
		\setlength{\parsep}{0pt}
		\setlength{\topsep}{0pt}
		\setlength{\partopsep}{0pt}
		\setlength{\leftmargin}{2em}
		\setlength{\labelwidth}{1.5em}
		\setlength{\labelsep}{0.5em}
} }

\newcommand{\spara}[1]{\smallskip\noindent\textbf{#1}}

\newcommand{\kw}[1]{{\ensuremath {\mathsf{#1}}}\xspace}
\newcommand{\nkw}[1]{{\ensuremath {\mathsf{#1}}}}

\newcommand{\stitle}[1]{\vspace{1ex} \noindent{\bf #1}}
\long\def\comment#1{}

\newtheorem{theorem}{Theorem}

\newcommand{\BD}{\kw{Online}-\kw{BCC}}
\newcommand{\BDQL}{\kw{LP}-\kw{BCC}}
\newcommand{\LBCC}{\nkw{L}$^2$\kw{P}-\kw{BCC}}
\newcommand{\CTC}{\kw{CTC}}
\newcommand{\PSA}{\kw{PSA}}
\newcommand{\Quarter}{Baidu-1\xspace}
\newcommand{\Year}{Baidu-2\xspace}
\newcommand{\Result}{O}

\newcommand{\MultipleLabels}{m}

\newcommand{\Nodenum}{|V|}
\newcommand{\Edgenum}{|E|}
\newcommand{\commonNei}[2]{N(#1)\cap N(#2)}
\newcommand{\lp}{\rho}
\newcommand{\cn}{\alpha}
\newcommand{\bcount}{\beta}

\usepackage{color}

\begin{document}
\title{Butterfly-Core Community Search over Labeled Graphs}

\author{Zheng Dong$^{1}$, Xin Huang$^{2}$, Guorui Yuan$^{1}$, Hengshu Zhu$^1$, Hui Xiong$^{3}$}
\affiliation{%
	\institution{$^1$Baidu Talent Intelligence Center, Baidu Inc.\\
		$^2$Hong Kong Baptist University \ $^3$Rutgers University}
}
\affiliation{\{dongzheng01, yuanguorui, zhuhengshu\}@baidu.com, xinhuang@comp.hkbu.edu.hk, xionghui@gmail.com}


\begin{abstract}
Community search aims at finding densely connected subgraphs for query vertices in  a graph. While this task has been studied widely in the literature, most of the existing works only focus on finding homogeneous communities rather than heterogeneous communities with different labels. In this paper, we motivate a new problem of cross-group community search, namely Butterfly-Core Community (BCC), over a labeled graph, where each vertex has a label indicating its properties and an edge between two vertices indicates their cross relationship. Specifically, for two query vertices with different labels, we aim to find a densely connected cross community that contains two query vertices and consists of butterfly networks, where each wing of the butterflies is induced by a k-core search based on one query vertex and two wings are connected by these butterflies. Indeed, the BCC structure admits the structure cohesiveness and minimum diameter, and thus can effectively capture the heterogeneous and concise collaborative team. Moreover, we theoretically prove this problem is NP-hard and analyze its non-approximability. To efficiently tackle the problem, we develop a heuristic algorithm, which first finds a BCC containing the query vertices, then iteratively removes the farthest vertices to the query vertices from the graph. The algorithm can achieve a $2$-approximation to the optimal solution. To further improve the efficiency, we design a butterfly-core index and develop a suite of efficient algorithms for butterfly-core identification and maintenance as vertices are eliminated. Extensive experiments on seven real-world networks and four novel case studies validate the effectiveness and efficiency of our algorithms.

\end{abstract}


\maketitle

\pagestyle{\vldbpagestyle}
\begingroup\small\noindent\raggedright\textbf{PVLDB Reference Format:}\\
\vldbauthors. \vldbtitle. PVLDB, \vldbvolume(\vldbissue): \vldbpages, \vldbyear.\\
\href{https://doi.org/\vldbdoi}{doi:\vldbdoi}
\endgroup
\begingroup
\renewcommand\thefootnote{}\footnote{\noindent
This work is licensed under the Creative Commons BY-NC-ND 4.0 International License. Visit \url{https://creativecommons.org/licenses/by-nc-nd/4.0/} to view a copy of this license. For any use beyond those covered by this license, obtain permission by emailing \href{mailto:info@vldb.org}{info@vldb.org}. Copyright is held by the owner/author(s). Publication rights licensed to the VLDB Endowment. \\
\raggedright Proceedings of the VLDB Endowment, Vol. \vldbvolume, No. \vldbissue\ %
ISSN 2150-8097. \\
\href{https://doi.org/\vldbdoi}{doi:\vldbdoi} \\
}\addtocounter{footnote}{-1}\endgroup

\ifdefempty{\vldbavailabilityurl}{}{
\vspace{.3cm}
\begingroup\small\noindent\raggedright\textbf{PVLDB Artifact Availability:}\\
The source code, data, and/or other artifacts have been made available at \url{\vldbavailabilityurl}.
\endgroup
}

\section{Introduction}\label{sec:intro}
Graphs are extensively used to represent real-life network data, e.g., social networks, academic collaboration networks, expertise networks, professional networks, and so on. Indeed, most of these networks can be regarded as labeled graphs, where vertices are usually associated with attributes as labels (e.g., roles in IT professional networks). A unique topological structure of labeled graph is the cross-group community, which refers to the subgraph formed by two knit-groups with close collaborations but different labels. For example, a cross-role business collaboration naturally forms a cross-group community in the IT professional networks. 

In the literature, numerous community models have been proposed for community search based on various kinds of dense subgraphs, e.g. quasi-clique~\citep{CuiXWLW13}, $k$-core~\citep{cui2014local, li2015influential, sozio2010}, $k$-truss~\citep{huang2015approximate}, and densest subgraph. For example, in the classical $k$-core based community model, a subgraph of $k$-core requires that each vertex has at least $k$ neighbors within $k$-core~\citep{batagelj2003m, seidman1983network}. The cohesive structure of $k$-core ensures that group members are densely connected with at least $k$ members. However, most of the existing studies only focus on finding homogeneous communities~\citep{fang2016effective, huang2017attribute}, which treat the semantics of all vertices and edges without differences. 


\begin{figure}[t]
\centering 
\includegraphics[width=0.8\linewidth]{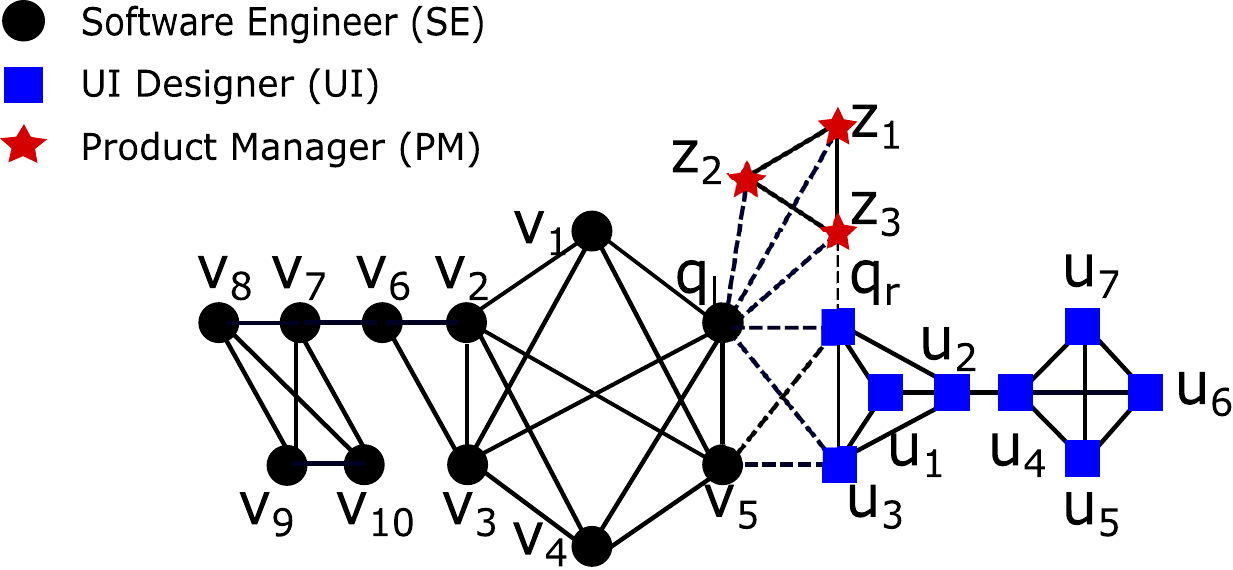}

\vspace*{-0.3cm}
\caption{An example of labeled graph $G$ in IT professional networks with three labels denote in different shapes and colors: SE, UI and PM. The collaborations between two employees of the same role (across over different roles) denote by the solid edges (dashed edges).}
\vspace*{-0.6cm}
\label{fig.intro}
\end{figure}


\vspace{-1mm}
\spara{Motivating example.} Figure~\ref{fig.intro} shows an example of IT professional networks $G$, where each vertex represents an employee and an edge represents the collaboration relationship between two employees. The vertices have three shapes and colors, which represent three different roles: ``Software Engineer (SE)'', ``UI Designer (UI)'', and ``Product Manager (PM)''. The edges have two types of solid and dashed lines. A solid edge represents a collaboration between two employees of the same role. A dashed edge represents a collaboration across over different roles, e.g., the dashed edge $(q_l, q_r)$ represents the collaboration between two employees of SE and UI roles respectively. Our motivation is how to effectively find communities formed by these cross-group collaborations given two employees with different roles.
Interestingly, considering a search for cross-group communities containing two query vertices $Q = \{q_l, q_r\}$, we find 
that conventional community search models cannot discover satisfactory results:

 \begin{list}{$\bullet$}
  { \setlength{\itemsep}{0pt}
    \setlength{\parsep}{0pt}
    \setlength{\topsep}{0pt}
    \setlength{\partopsep}{0pt}
    \setlength{\leftmargin}{2em}
    \setlength{\labelwidth}{1.5em}
    \setlength{\labelsep}{0.5em}
}

\item \spara{Structural community search}. This kind algorithms find communities containing all query vertices over a simple graph, which ignores the vertex labels and treats $G$ as a homogeneous graph. W.l.o.g. we select $k$-core~\citep{seidman1983network} as an example, the maximum core value of $q_l, q_r$ are $4$ and $3$ respectively, one limitation of this model is the smaller vertex coreness dominates $k$ value to contain all query vertices. Each vertex on $G$ has a degree of at least $3$, thus the whole graph of $G$ is returned as the answer. However, the model suffers from several disadvantages: (1) it fails to capture different community densities of two teams. (2) it treats the semantics of all edges equally, which not only ignores the semantics of different edges but also mixes different teams.
(3) the vertices span a long distance to others, e.g., the distance between $v_8$ and $u_{7}$ is 8. Many vertices are irrelevant to the query vertices, such as the vertices $\{v_6, v_7, v_8, v_9, v_{10}\}$ and $\{u_4, u_5, u_6, u_7\}$. (4) the vertices with irrelevant labels to the query vertices, e.g., $z_1$'s label is PM different with SE and UI. Other community search models add graph size constraints such as the  minimum size of $k$-core~\citep{li2019efficient} or the minimum diameter~\citep{huang2015approximate}. However, such improved models find the answer of $\{q_l, q_r, v_5, u_3\}$, which suffers from missing many group members with no cross-group edges. 



\item \spara{Attributed community search}. The studies of attributed community~\citep{fang2016effective, huang2017attribute} focus on identifying the communities that have cohesive structure and share homogeneous labels. For instance, using query vertices $Q = \{q_l, q_r\}$, keywords $W_q = \{SE, UI\}$ as input, \citep{fang2016effective} returns a $k$-core subgraph and maximizes the number of keywords all the vertices share. Since the vertices only contain one label (keyword) on the labeled graph, the cross-group community share no common attributes, then the keyword cohesiveness is always $0$, it will return empty result. 




\begin{figure}[t]
\centering 
\includegraphics[width=0.5\linewidth]{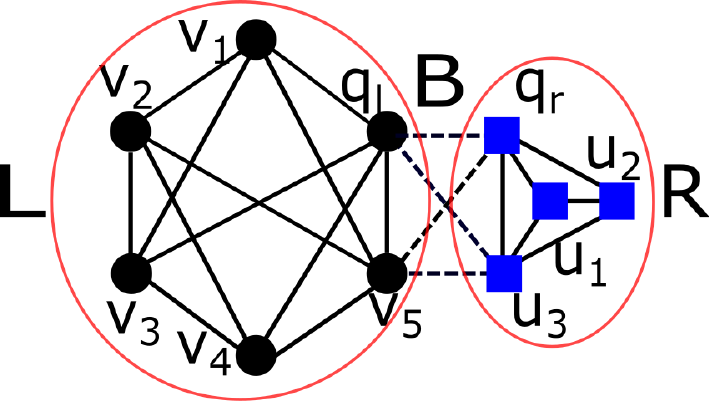}
\vspace*{-0.4cm}
\caption{An example of butterfly-core community $H$ on $G$ in Figure~\ref{fig.intro}. 
$B$ is a bow tie formed by all dashed edges across two labeled groups.
}
\vspace*{-0.6cm}
\label{fig.intro-bcc}
\end{figure}

\item \spara{Ours}. The expected answer of our proposed \emph{Butterfly-Core Community (BCC) search}, aims to find cross-group communities using two query vertices, as shown in Figure~\ref{fig.intro-bcc}.
The cross-group community has three key parts. The first $L$ is the induced subgraph formed by the vertices with the label SE, which is a $4$-core. The second $R$ is the $3$-core of vertices with the label UI. The third part $B$ is the bipartite graph (subgraph induced by vertices $\{q_l, q_r, v_5, u_3\}$ with only dashed edges) across over two groups of SE and UI vertices containing a butterfly, i.e., a complete $2 \times 2$ biclique.

\end{list}


Motivated by the above example, in this paper, we study a novel problem of cross-group community search in the labeled graph, namely BCC Search. Specifically, given a labeled graph $G$, two query vertices with different labels $Q = \{q_l, q_r\} \in G$ and integers $\{k_1, k_2, b\}$, we define the $(k_1, k_2, b)$-BCC search problem as to find out a densely connected cross-group community that is related to query vertices $Q$.





In light of the above, we are interested in developing efficient algorithms for the BCC search problem. However, the efficient extraction of BCCs raises significant challenges. We theoretically prove that the BCC search problem is NP-hard and cannot be approximated in polynomial time within a factor $(2 - \varepsilon)$ of an optimal answer with the smallest diameter for any small $\varepsilon \in (0, 1)$ unless $P = NP$. Therefore, we develop a greedy algorithmic framework, which first finds a BCC containing $Q$ and then iteratively maintains BCC by removing the farthest vertices to $Q$ from the graph. The method can achieve a $2$-approximation to the optimal BCC answer, obtaining no greater than twice the smallest diameter. 
To further improve efficiency, we construct the offline butterfly-core index and develop efficient algorithms for butterfly-core identification and maintenance. 
In addition, we further develop a fast algorithm \LBCC, which integrates several optimization strategies including the bulk deletion of removing multiple vertices each time, the fast query distance computation, leader pair strategy, and the local exploration to generate a small candidate graph. We further discuss how to extend the BCC model to handle queries with multiple vertex labels. To summarize, we make the following contributions. 


\begin{list}{$\bullet$}
  { \setlength{\itemsep}{0pt}
    \setlength{\parsep}{0pt}
    \setlength{\topsep}{0pt}
    \setlength{\partopsep}{0pt}
    \setlength{\leftmargin}{2em}
    \setlength{\labelwidth}{1.5em}
    \setlength{\labelsep}{0.5em}
}
\item We study and formulate a novel problem of BCC search over labeled graphs. We propose a $(k_1, k_2, b)$-BCC model to find a cross-group community containing two query vertices $q_l$ and $q_r$ with different labels. Moreover, we give the BCC-problem analysis and illustrate useful applications.  (Section~\ref{sec:butterfly}).

\item We show the BCC problem is NP-hard and cannot be approximated in polynomial time within a factor $(2 - \varepsilon)$ of the optimal diameter for any small $\varepsilon \in (0, 1)$ unless $P = NP$ (Section~\ref{sec:hardness}).

\item We develop a greedy algorithm for finding BCC containing two query vertices, which achieves a $2$-approximation to an optimal BCC answer with the smallest diameter. The algorithm iteratively deletes the farthest vertices from a BCC, which achieves a small diameter (Section~\ref{sec:algorithm}).

\item We develop several improved strategies for fast BCC search. 
First, we design an efficient bulk deletion strategy to remove multiple vertices at each iteration; Second, we optimize the shortest path computations of two query vertices; Third, we make a leader pair algorithm for butterfly count maintenance; Finally, we propose an index based local search method (Section~\ref{sec:fast}). 

\item
We extend the BCC model to handle cross-group communities with multiple vertex labels and leverage our \LBCC techniques to develop an efficient search solution (Section~\ref{sec:multiple}).

\item We conduct extensive experiments on seven real-world datasets with ground-truth communities.
Four interesting case studies of cross-group communities are discovered by our BCC model on real-world \emph{global flight networks}, \emph{international trade networks}, \emph{complex fiction networks}, and \emph{academic collaboration networks}. The results show that our proposed algorithms can efficiently and effectively discover BCC, which significantly outperform other approaches (Section~\ref{sec:exp}).
\end{list}

We discuss related work in Section~\ref{sec:related}, and conclude the paper with a summary in Section~\ref{sec:con}.


\section{Related Work}\label{sec:related}


\spara{Attributed community discovery}. 
The studies of attributed community discovery involve two problems of attributed community detection and attributed community search. Attributed community detection is to find all communities in an attributed graph where vertices have attributed labels~\citep{bothorel2015clustering, zhou2009graph, guo2021multi, luo2020efficient}. Thus, attributed community detection is not the same as our problem in terms of community properties and input data. A survey of clustering on attributed graphs can be found in~\citep{bothorel2015clustering}.
In addition, given a set of query vertices and query attributes, attributed community search finds the query-dependent communities in which vertices share homogeneous query attributes~\citep{fang2016effective, huang2017attribute, vac2020liu}. Most recently, Zhang et al.~\citep{zhang2019keyword} proposed an attributed community search model using only query keywords but no query vertices. 
Other related works to ours are community detection in heterogeneous networks \citep{sun2013mining} where vertices have various vertex labels. However, heterogeneous communities are defined based on meta patterns, which are different from our communities across over two labeled groups. Compared with all the above studies, our butterfly-core community search is a novel problem over labeled graphs, which has not been studied before. 
 


\spara{Community search}. Community search finds the query-dependent communities in a graph~\citep{huang2019community, fang2019survey, chen2019contextual}. 
Community search models can be categorized based on different dense subgraphs including $k$-core~\citep{cui2014local, li2015influential, barbieri2015efficient, sozio2010, lin2021hierarchical, wang2019forbidden}, $k$-truss~\citep{huang2015approximate}, quasi-clique~\citep{CuiXWLW13}, and densest subgraph~\citep{wu2015robust}. Sozio and Gionis defined the problem of community search and proposed a $k$-core based model with the distance and size constraints~\citep{sozio2010}. All these community models work on simple structural graphs, which ignore the vertex labels. Recently, several complex community models have been studied for various graph data, such as 
directed graphs~\citep{truss2020liu, li2017most, fang2018effective}, weighted graphs~\citep{sun2019fast, zheng2017querying}, spatial-social networks~\citep{kim2020densely, al2020topic, chen2020finding, chen2018maximum} and so on. Most attributed community search studies aim at finding the communities that have a dense structure and similar attributes~\citep{fang2016effective, huang2017attribute, zhang2019keyword, vac2020liu}. This is different from our studies over labeled graphs, which trends to find two groups with different labels. 
Our community model requires the dense structure to appear not only in the inter-groups but also between two intra-groups.
Most recently, there are two studies that investigate community search on heterogeneous information networks~\citep{jian13effective, fang2020effective}, where vertices belong to multiple labeled types. Fang et al.~\citep{fang2020effective} leveraged meta-path patterns to find communities where all vertices have the same label of a query vertex and close relationships following the given meta-paths. Jian et al.~\citep{jian13effective} proposed the relational constraint to require connections between labeled vertices in a community. They developed heuristic solutions for detecting and searching relational communities due to the hard-to-approximate problem. Both studies are different from our BCC search model that takes two query vertices with different labels and finds a leader pair based community integrating two cross-over groups. Our problem is NP-hard but can be approximately tackled in polynomial time with an approximate ratio of 2.

\spara{Butterfly counting}.
In the bipartite graph analytics~\citep{wang2020efficient, sariyuce2018peeling}, the butterfly is a cohesive structure of $2\times2$ biclique. 
Butterfly counting is to calculate the number of butterflies in a bipartite graph~\citep{sanei2018butterfly, wang2019vertex, sanei2019fleet, li2021approximately, yang2021efficient}. Sanei et al.~\citep{sanei2018butterfly} proposed exact butterfly counting and approximation solutions using randomized strategies. Wang et al.~\citep{wang2019vertex} further optimized the butterfly counting by assigning high degree vertex with high priority to visit wedges.
However, these studies focus on the \emph{global butterfly counting} to compute the butterfly number over an entire graph.
While our BCC search algorithms aim at finding a few vertices with large butterfly degrees as the leaders of cross-group communities. Moreover, our algorithms can dynamically update such leader vertices to admit butterfly degrees when the graph structure changes. Overall, our proposed butterfly search solutions are efficient to find leader vertices and update butterfly degrees \emph{locally}, which can avoid the \emph{global} butterfly counting multiple times. 
\section{Problem Formulation}\label{sec:butterfly}

In this section, we  introduce the definition and our problem.

\vspace{-2mm}
\subsection{Labeled Graph}

Let $G = (V, E, \ell)$ be a labeled graph, where $V$ is a set of vertices, $E \subseteq V\times V$ is a set of undirected edges, and $\ell:V \rightarrow \mathcal{A}$ is a vertex label function mapping from vertices $V$ to labels $\mathcal{A}$. For each vertex $v\in V$, $v$ is associated with a label $\ell(v)\in \mathcal{A}$. The edges have two types, i.e., given two vertices $\{u, v\} \in V$, if with same label $\ell(u)=\ell(v)$, $(u, v)$ is a homogeneous edge; otherwise, if $\ell(u)\neq \ell(v)$, $(u, v)$ is a heterogeneous (cross) edge. For example, consider $G$ in Figure~\ref{fig.intro}, $G$ has three labels: SE, UI and PM. The vertex $v_1$ has a label of SE. The edge $(v_1, v_2)$ is a homogeneous edge for (SE-SE). The edge $(v_5, u_3)$ is a heterogeneous edge for (SE-UI). 

Given a subgraph $H\subseteq G$, the degree of a vertex $v$ in $H$ is denoted as $deg_H(v) = |N_H(v)|$, where $N_H(v)$ is the set of $v$'s neighbors in $H$. For two vertices $u, v$, we denote $dist_H(u, v)$ as a length of the shortest path between $u$ and $v$ in $H$, where  $dist_H(u, v) = \infty$ if $u$ and $v$ are disconnected. The diameter of $H$ is defined as the maximum length of the shortest path in $H$, i.e., $diam(H) = \max_{u,v \in V(H)}{dist_H(u, v)}$~\citep{huang2015approximate}.

\begin{table}[t]
	\centering
	\small
	\caption{Frequently used notations.}\label{tab:symbols}
	\vspace{-0.3cm}
	\begin{tabular}{l|l }
		\hline
		{\bf Notation} & {\bf Description} \\ \hline
		$G = (V, E, \ell)$  & a labeled graph $G$ with a vertex label function $\ell$ \\
		$q_l, q_r$ & query vertices $q_l$, $q_r$  \\
		$L, B, R$ & left $k_1$-core, bipartite graph, right $k_2$-core  \\
		$V_L, V_R$ & the vertex sets of left $k_1$-core and right $k_2$-core \\
		$dist_H(u, v)$ & the length of the shortest path between $u$ and $v$ in $H$ \\
		$diam(H)$ & the diameter of a graph $H$ \\
		$\chi(v)$ & the butterfly degree of vertex $v$ \\
		$\ell(v)$ & the label associated with vertex $v$  \\ 
		$N(v)$ & the set of neighbor vertices of vertex $v$ in graph $G$  \\ 
		$\commonNei{u}{v}$ & the common neighbors of $u$ and $v$ \\
		
		$N_v^2$ & the vertices within $v$'s 2-hop neighborhood (excluding $v$)  \\ \hline
		

	\end{tabular}
	\vspace{-0.4cm}
\end{table}



\vspace{-2mm}
\subsection{K-Core and Butterfly}
\vspace{-0.5mm}

We give two definitions of $k$-core~\citep{seidman1983network} and butterfly~\citep{sanei2018butterfly, wang2019vertex}. 
\begin{definition}[$k$-core]
Given a subgraph $H\subseteq G$ and an integer $k$, $H$ is a $k$-core if each vertex $v$ has at least $k$ neighbors within $H$, i.e., $deg_{H}(v) \geq k$.
\end{definition}

The coreness $\delta(v)$ of a vertex $v\in V$ is defined as the largest number $k$ such that there exists a connected $k$-core containing $v$. In Figure~\ref{fig.intro-bcc}, $L$ is a $4$-core as each vertex has at least $4$ neighbors within $L$. 
Next, we define the butterfly~\citep{sanei2018butterfly, wang2019vertex} in a bipartite graph.


\begin{definition}[Butterfly]
Given a bipartite graph $B = (V_L, V_R, E)$ where $E\subseteq V_L \times V_R$, a butterfly $H$ is a $2 \times 2$ biclique of $G$ induced by four vertices 
 $v_{l_1}, v_{l_2} \in V_L$, $v_{r_1}, v_{r_2} \in V_R$ such that all four edges $(v_{l_1}, v_{r_1})$, $(v_{l_1}, v_{r_2})$, $(v_{l_2}, v_{r_1})$ and $(v_{l_2}, v_{r_2})$ exist in $H$.
\end{definition}

\begin{definition}[Butterfly Degree]
Given a bipartite graph $B = (V_L, V_R, E)$, the butterfly degree of vertex $v$ is the number of butterfly subgraphs containing $v$ in $G$, denoted by $\chi(v)$. 
\end{definition}

\begin{example}
In Figure~\ref{fig.intro-bcc}, the subgraph $B$ is a butterfly since it is a $2 \times 2$ biclique formed by four vertices $\{q_l, v_5, q_r, u_3\}$. There exists a unique butterfly $B$ containing the vertex $q_r$. Thus, the butterfly degree of $q_r$ is  $\chi(q_r) =1$.
\end{example}

\subsection{Butterfly-Core Community Model}
We next discuss a few choices to model the cross-group  relationships between two groups $V_L$ and $V_R$ with different labels in the community $H$, and analyze their pros and cons. To quantify the strength of cross-group connections, we use the number of butterflies between two groups, denoted as $b$. 

\begin{list}{$\bullet$}
  { \setlength{\itemsep}{0pt}
    \setlength{\parsep}{0pt}
    \setlength{\topsep}{0pt}
    \setlength{\partopsep}{0pt}
    \setlength{\leftmargin}{2em}
    \setlength{\labelwidth}{1.5em}
    \setlength{\labelsep}{0.5em}
}
\item First, we consider that $\chi(v) \geq b$ for each vertex $v\in H$. It requires that each vertex's butterfly count is at least $b$, i.e., $\chi(v) \geq b$. This constraint is too strict, which may miss some vertices without heterogeneous edges. Take Figure~\ref{fig.intro} as an example, some vertices act like leaders or liaisons who are in charge of communications across the groups, i.e., $\{q_l, q_r, v_5, u_3\}$, while some vertices mostly link within their own group with less interactions across the groups such as $\{v_1, v_2, v_3, v_4\}$. If we model in this way, an input $Q = \{v_1, q_r\}$ requires that $v_1$ exists in at least one butterfly, which is impossible.

\item Second, we alternatively consider $\sum_{v\in V(H)}\chi(v) \geq b$, which requires the total butterfly count in $H$ is at least $\lceil b/4 \rceil$. However, it is hard for us to determine the parameter $b$ as we cannot estimate a qualified number of butterflies in community $H$, which is a global criterion varying significantly over different kinds of graphs.

\item Finally, we consider a constraint between two groups of vertices $V_L$ and $V_R$ that $\exists v_l \in V_L$ and $\exists v_r \in V_R$ to make $\chi(v_l) \geq b$ and $\chi(v_r) \geq b$ hold. It is motivated by real applications. Generally, one collaboration community has at least one leader or liaison in each group, so we require there exists at least one vertex in each group whose butterfly count is at least $b$. 
In this setting, no matter the input query vertices are leaders biased (e.g., $Q = \{q_l, q_r\}$) or juniors biased (e.g., $Q = \{v_1, u_1\}$), the underlying community is identical. 
\end{list}


In view of these considerations, we define the butterfly-core community as follows.

\begin{definition}[Butterfly-Core Community]
\label{def.bcc}
Given a labeled graph $G$, a $(k_1, k_2, b)$-butterfly-core community (BCC) $H \subseteq G$ satisfies the following conditions:

1. Two labels: there exist two labels $A_l, A_r\in \mathcal{A}$, $V_L =\{v\in H: \ell(v)=A_l\}$ and $V_R=\{v\in H: \ell(v)=A_r\}$ such that $V_L\cap V_R=\emptyset$ and $V_L\cup V_R = V_H$;

2. Left core: the induced subgraph of $H$ by $V_L$ is $k_1$-core; 

3. Right core: the induced subgraph of $H$ by $V_R$ is $k_2$-core; 

4. Cross-group interactions: 
$\exists v_l \in V_L$ and $\exists v_r\in V_R$ such that butterfly degree $\chi(v_l) \geq b$ and $\chi(v_r) \geq b$ hold.

\end{definition}

In terms of vertex labels, condition (1) requires that the BCC contains exactly two labels for all vertices. In terms of homogeneous groups, conditions (2) and (3) ensure that each homogeneous group satisfies the cohesive structure of $k$-core, in which community members are internally densely connected. In terms of cross-group interactions, condition (4) targets two representative vertices of two homogeneous groups, which have a required number of butterflies with densely cross-group interactions.
Moreover, we call the two vertices $v_l$ and $v_r$ with $\chi(v_l) \geq b$ and $\chi(v_r) \geq b$ as a \emph{leader pair}. 

\begin{example}
Figure~\ref{fig.intro-bcc} shows a $(4, 3, 1)$-BCC. The subgraphs $L$ and $R$ are respectively the left 4-core group and the right 3-core group, respectively. The subgraph $B$ is a butterfly across over two groups $L$ and $R$,  and $\chi(q_l) =\chi(q_r)=1$.
\end{example}

\subsection{Problem Formulation}

We formulate the BCC-Problem studied in this paper. 

\vspace{-1mm}
\begin{problem}\label{def.bcc_problem}
(BCC-Problem) Given $G(V, E, \ell)$, two query vertices $Q = \{q_l, q_r\} \subseteq V$ and three integers $\{k_1, k_2, b\}$, the BCC-Problem finds a BCC $H\subseteq G$, such that:

1. Participation \& Connectivity: $H$ is a connected subgraph  containing $Q$;

2. Cohesiveness: $H$ is a $(k_1, k_2, b)$-BCC.



3. Smallest diameter: $H$ has the smallest diameter, i.e., $ \nexists H' \subseteq G $, such that $diam(H') < diam(H)$, and $H'$ satisfies the above conditions $1$ and $2$.
\end{problem}
\vspace{-1mm}

The BCC-Problem prefers a tight BCC with the smallest diameter such that group members have a small communication cost, to remove query unrelated vertices.
In addition, we further study the BCC-problem for multiple query vertices, generalizing the BCC model from $2$ group labels to $m$ group labels where $m\geq2$ in Section~\ref{sec:multiple}.

\begin{example}
Consider a labeled graph $G$ in Figure~\ref{fig.intro}. Assume that the inputs $Q = \{q_l, q_r\}$, $k_1 = 4$, $k_2 = 3$, and $b = 1$. The answer is the $(4, 3, 1)$-butterfly-core community containing $Q$ as shown in Figure~\ref{fig.intro-bcc}.
\end{example}

\vspace{-1mm}
\subsection{Why Butterfly-Core Community Model?}
\spara{Why butterfly}. A butterfly is a complete bipartite subgraph of $2\times2$ vertices, which serves as the fundamental motif in bipartite graphs. For two groups $V_L$ and $V_R$ with different labels, we model the collaborative interactions between two groups $V_L$ and $V_R$ using the butterfly model~\citep{borgatti1997network, robins2004small, derr2019balance}. 
More butterflies indicate 1) stronger connections between two groups and 2) similar properties sharing within the same group members, which are validated in many application scenarios. For instance, in the users-items bipartite graph $G$, a user $x_1 \in V_L$ buy an item $y_1\in V_R$, then we have an edge $(x_1, y_1)$ in $G$. Thus, two users $x_1$ and $x_2$ buy the same two items $y_1$ and $y_2$, which forms a butterfly in $G$. Two users purchase the same items, indicating the more similar purchasing preferences of them and more butterflies in the community. Similar cases happen in the common members of \emph{the board of directors} between \emph{two different companies} and also the common members of \emph{the steering committee} in \emph{two conference organizations}. Moreover, in the email communication networks, threads of emails are delivered between \emph{two partner groups} and also cc's to superiors on both sides. The superiors of two partner groups receive the most emails and play the \emph{leader pair} positions of our BCC model. Overall, the butterfly plays an essential role as the basis higher-order motif in bipartite graphs, which can be regarded as an implicit connection measure between two same labeled vertices. 

\spara{Why BCC model}. The BCC model inherits several good structural properties and efficient computations. First, the community structure enjoys \underline{high computational efficiency}. The $k$-core is a natural and cohesive subgraph model of communities in real applications, requiring that every person has at least $k$ neighbors in social groups, which can be computed faster than $k$-clique. In addition, butterfly listing takes a polynomial time complexity and enjoys an efficient enumeration, which could be optimized by assigning the wedge visiting priority based on vertex degrees~\citep{robins2004small, wang2019vertex}. Second, two labeled groups in BCC model \underline{admit practical cases of different group densities} in real-world applications. Our BCC model crosses over two labeled groups, which may have different group sizes and densities. Thus, two different $k$-core parameters, i.e., $k_1$ and $k_2$ are greatly helpful to capture different community structures of two groups. One simple way for parameter setting is to automatically set $k_1$ and $k_2$ with the coreness of two queries $q_l$ and $q_r$ respectively. Third, \underline{automatic identification of leader pair} in the BCC discovery. The constraint of cross-group interactions is motivated by real-world scenarios that leaders or liaisons in each group always take most interactions with the other group. 

\subsection{Applications}
In the following, we illustrate representative applications of butterfly-core community search.
\begin{list}{$\bullet$}
  { \setlength{\itemsep}{0pt}
    \setlength{\parsep}{0pt}
    \setlength{\topsep}{0pt}
    \setlength{\partopsep}{0pt}
    \setlength{\leftmargin}{2em}
    \setlength{\labelwidth}{1.5em}
    \setlength{\labelsep}{0.5em}
}
\item  \textbf{Interdisciplinary collaboration search}. Given two principal investigators from different departments in the universities, who intend to form a team to apply for an interdisciplinary research grant. The team is better formed by two cohesive groups with good inner-group communications. Moreover, the principal investigators or liaisons should also have cross-group communications. 
\item \textbf{Professional team discovery}. In high-tech companies, there are usually many cross-department projects between two teams with different sizes of employees. Moreover, the technical leader and product manager of each team always take charge of the cross-group communications and information sharing, which naturally form a butterfly, i.e., $2 \times 2$ biclique. 
\item \textbf{Various real-world cross-group mining tasks}. BCC search can be applied on various real-world labeled graphs, e.g., \emph{global flight networks}, \emph{international trade networks}, \emph{complex fiction networks}, and \emph{academic collaboration networks}, as reported in four interesting case studies in Section~\ref{sec:exp}. 
\end{list}

\section{Hardness and Approximation}\label{sec:hardness}
In this section, we analyze the hardness and non-approximability of the BCC-Problem. 

\spara{Hardness}. 
We define a decision version of the BCC-Problem.

\begin{problem}
 (BCC-Decision Problem) Given a labeled graph $G(V, E, \ell)$, two query vertices $Q = \{q_l, q_r\} \subseteq V$ with different labels, 
 and parameters $\{k_1, k_2, b\}$, test whether $G$ has a connected butterfly-core subgraph containing $Q$ with a diameter at most $d$.
\end{problem}
\vspace{-2mm}

\vspace{-1mm}
\begin{theorem}
The BCC-Problem is NP-hard.
\end{theorem}

\vspace{-3mm}
\begin{proof}
We reduce the well-known NP-hard problem of Maximum Clique (decision version) to BCC-Problem. Given a graph $G(V, E)$ and a number $k$, the maximum clique decision problem is to check whether $G$ contains a clique of size $k$. From this, construct an instance $G^{\prime}(V^{\prime}, E{^\prime}, \ell)$ of BCC-Problem as follows. $G^{\prime} = (V^{\prime}:V + V_c, E^{\prime}:E + E_c + E_b, \ell)$. For each vertex $v \in V$ we assign the label of $l_1$, i.e., $\ell(v) = l_1, \forall v \in V$. $G_c(V_c, E_c)$ is a copy of $G(V, E)$ associated with labels $l_2$, i.e., $\ell(v) = l_2, \forall v \in V_c$. $E_b$ is the edge set that connects any two vertices $v_i$ and $v_j$ where $v_i \in G$ and $v_j \in G_c$, i.e., $E_b = V \times V_c$. Set parameters $k_1 = k_2 = k-1$, $b = 1$ (actually $b$ could choose any value fits $b \leq (k-1) \times {k \choose 2}$), $d = 1$ and the query vertices $Q = \{q_l, q_r\}$ where $q_l \in G$ and $q_r \in G_c$, i.e., $\ell(q_l) \neq \ell(q_r)$. We show that the instance of the maximum clique decision problem is a YES-instance iff the corresponding instance of BCC-Problem is a YES-instance.

\vspace{-0mm}
$(\Rightarrow):$ Clearly, any clique with at least $k$ vertices is a connected $(k-1)$-core, since $G_c$ is a copy of $G$ then there will be a connected $(k-1)$-core in $G_c$. Because any edge between the vertices in $G$ and $G_c$ are connected then for each vertex $v \in V$, it forms $1$ butterfly with any vertex exclude itself in $V$ and any two vertices in $V_c$; the same proof to the vertex $v \in V_c$. Then $H$ is a $(k-1, k-1, 1)$-BCC with a diameter $1$.

\vspace{-0mm}
$(\Leftarrow):$ Given a solution $H$ for BCC-Problem, we split $H$ into two parts $H_l$ whose vertices label is $\ell(q_l)$ and $H_r$ whose vertices label is $\ell(q_r)$. Since $H$ is a $(k-1, k-1, 1)$-BCC, $H_l$ and $H_r$ must contain at least $k$ vertices, $diam(H) = d = 1$ implies $H_l$ and $H_r$ are both cliques which implies $G$ has a clique since $G_c$ is a copy of $G$.
\vspace{-0mm}
\end{proof}

Given the NP-hardness of the BCC-Problem, it is interesting whether it can be approximately tackled. 
We analyze the approximation and non-approximability as follows.

\spara{Approximation and non-approximability}. For $\alpha \geq 1$, we say that an algorithm achieves an $\alpha$-approximation to  BCC-Problem if it outputs a connected $(k_1, k_2, b)$-BCC $H \subseteq G$ such that $Q \subseteq H$ and $diam(H) \leq \alpha \cdot diam(H^\ast)$, where $H^\ast$ is the optimal BCC. That is, $H^\ast$ is a connected $(k_1, k_2, b)$-BCC 
s.t. $Q \subseteq H^\ast$, and diam($H^\ast$) is the minimum among all such BCCs containing $Q$. 


\begin{theorem}
Unless $P = NP$, for any small $\epsilon \in (0, 1)$ and given parameters $\{k_1, k_2, b\}$,  BCC-Problem cannot be approximated in polynomial time within a factor $(2-\varepsilon)$ of the optimal.
\end{theorem}

\vspace{-2mm}
\begin{proof}
We prove it by contradiction. Assume that there exists a $(2-\varepsilon)$-approximation algorithm for the BCC-Problem in polynomial time complexity, no matter how small the $\epsilon \in (0, 1)$ is. This algorithm can distinguish between the YES and NO instances of the maximum clique decision problem. That is, if an approximate answer of the reduction problem has a diameter of 1, it corresponds to the Yes-instance of maximum clique decision problem; otherwise, the answer with a diameter value of no less than 2 corresponds to the No-instance of the maximum clique decision problem. This is impossible unless P=NP.
\end{proof}



\section{BCC Online Search Algorithms}\label{sec:algorithm}

\begin{algorithm}[t]
	\small
	\caption{BCC Online Search $(G, Q)$} \label{algo:Basic}
	\begin{algorithmic}[1]
		\REQUIRE $G = (V, E, \ell)$, $Q = \{q_l, q_r\}$, three integers $\{k_1, k_2, b\}$.
		\ENSURE A connected $(k_1, k_2, b)$-BCC $\Result$ with a small diameter.
		\STATE Find a maximal connected $(k_1, k_2, b)$-BCC containing $Q$ as $G_0$; //see Algorithm~\ref{algo:bccs} \label{line:a1}
		\STATE $l \gets 0$; \label{line:a2}
		
		\WHILE{$connect_{G_l} (Q) = \textbf{true}$} \label{line:a3}
		\STATE Compute dist$_{G_l} (q, u), \forall q \in Q$ and $\forall u \in G_l$; \label{line:a4}
		\STATE  $u^* \gets argmax_{u \in G_l} dist_{G_l} (u, Q)$; \label{line:a5}
		\STATE  $dist_{G_l} (G_l, Q) \gets dist_{G_l} (u^*, Q)$; \label{line:a6}
		\STATE  Delete $u^*$ and its incident edges from $G_l$; \label{line:a7}
		\STATE  Maintain $G_l$ as a $(k_1, k_2, b)$-BCC; //see Algorithm~\ref{algo:maintenance} \label{line:a8}
		\STATE  $G_{l+1} \gets G_l$; $l \gets l + 1$; \label{line:a9}
		\ENDWHILE
		\STATE $\Result \gets \mathop{\arg\min}_{G' \in \{G_0, ..., G_{l-1}\}} dist_{G'}(G', Q)$; \label{line:a10}
	\end{algorithmic}
\end{algorithm}

In this section, we present a greedy algorithm for the BCC-problem, which online searches a BCC. Then, we show that the greedy algorithm can achieve a $2$-approximation to optimal answers. Finally, we discuss an efficient implementation of the algorithm and analyze the time and space complexity.

\subsection{BCC Online Search Algorithm}
We begin with a definition of query distance as follows.
\begin{definition}[Query Distance]
Given a graph $G$, a query set $Q$, and a set of vertices $X$, the query distance of $X$ is the maximum length of the shortest path from $v\in X$ to a query vertex $q \in Q$, i.e., $dist_G(X, Q) = max_{v\in X, q \in Q} dist_G(v, q)$. 
\end{definition}

For simplicity, we use $dist_G(H, Q)$ and $dist_G(v, Q)$ to represent the query distance for the vertex set $V_H$ in $H\subseteq G$ and a vertex $v\in V$. Motivated by \citep{huang2015approximate}, we develop a greedy algorithm to find a BCC with the smallest diameter. 
Here is an overview of the algorithm. First, it finds a maximal connected $(k_1, k_2, b)$-BCC containing $Q = \{q_l, q_r\}$, denoted as $G_0$. As the diameter of $G_0$ may be large, it then iteratively removes  from $G_0$ the vertices far away to $Q$, meanwhile it maintains the remaining graph as a $(k_1, k_2, b)$-BCC.

\spara{Algorithm.}
Algorithm~\ref{algo:Basic} outlines a greedy algorithmic framework for finding a BCC. The algorithm first finds $G_0$ that is a maximal connected $(k_1, k_2, b)$-BCC containing $Q$ (line \ref{line:a1}). Then, we set $l = 0$. For all $u \in G_l$ and $q \in Q$, we compute the shortest distance between $q$ and $u$, and obtain the vertex query distance $dist_{G_l}(u, Q)$ (line \ref{line:a4}). Among all vertices, we pick up a vertex $u^*$ with the maximum distance $dist_{G_l}(u^*, Q)$, which equals $dist_{G_l}(G_l, Q)$ (lines \ref{line:a5}-\ref{line:a6}). Next, we remove the vertex $u^*$ and its incident edges from $G_l$ and also delete vertices/edges to maintain $G_l$ as a $(k_1, k_2, b)$-BCC (lines \ref{line:a7}-\ref{line:a8}). 
Then, we repeat the above steps until $G_l$ is disqualified to be a BCC containing $Q$ (lines \ref{line:a3}-\ref{line:a9}). Finally, the algorithm terminates and returns a BCC $\Result$, where $\Result$ is one of the graphs $G' \in \{G_0,...,G_{l-1}\}$ with the smallest query distance $dist_{G'} (G', Q)$ (line \ref{line:a10}). Note that each intermediate graph $G' \in \{G_0,...,G_{l-1}\}$ is a $(k_1, k_2, b)$-BCC.



\subsection{Butterfly-Core Discovery and Maintenance}
\vspace{-1mm}
We present two important procedures for BCC online search algorithm: finding $G_0$ (line \ref{line:a1} in Algorithm \ref{algo:Basic}) and butterfly-core maintenance (line \ref{line:a8} in Algorithm \ref{algo:Basic}). 


\begin{algorithm}[t]
	\small
	\caption{Find $G_0 (G, Q)$} \label{algo:bccs}
	\begin{algorithmic}[1]
		\REQUIRE  $G = (V, E, \ell)$, $Q = \{q_l, q_r\}$, three integers $\{k_1, k_2, b\}$.
		\ENSURE A connected $\{k_1, k_2, b\}$-BCC $G_0$ containing $Q$.
		\STATE $V_L \gets \{v\in V\ |\ \ell(v) = \ell(q_l) \}$; $V_R \gets \{v\in V\ |\ \ell(v) = \ell(q_r) \}$;
		\STATE Let $L$ be a $k_1$-core induced subgraph of $G$ by $V_L$;
		\STATE Let $R$ be a $k_2$-core induced subgraph of $G$ by $V_R$;
		\STATE $B = \{V_B, E_B\}$, where $V_B = V_L \cup V_R$ and $E_B = \{V_L \times V_R\} \cap E$;
		\STATE Butterfly Counting$(B)$; // See Algorithm \ref{algo:count}
		\STATE $max_l \gets \max_{u \in V_L} \chi(u)$;
		\STATE $max_r \gets \max_{u \in V_R} \chi(u)$;
		\IF{$max_l < b$ \OR $max_r < b$}
		\RETURN $\emptyset$;
		\ENDIF
		\STATE $G_0 \gets L \cup B \cup R$;
	\end{algorithmic}
\end{algorithm}

\subsubsection{Finding $G_0$} As an essentially important step, finding $G_0$ is to identify a maximal connected $(k_1, k_2, b)$-BCC containing $Q$ in graph $G$. The challenge lies in finding a butterfly-core structure, which needs to shrink the graph by vertex removals. However, deleting vertices may trigger off the change of vertex coreness and butterfly degree for vertices in the remaining graph. 
To address it, our algorithm runs the $k$-core decomposition algorithm twice and then runs the  butterfly counting method once. 
The \emph{general idea} 
is to first identify a candidate subgraph formed by two groups of vertices sharing the same labels with $q_l$ and $q_r$. Then, it shrinks the graph by applying core decomposition algorithm, which deletes disqualified vertices to identify $k_1$-core and $k_2$-core, denoted by $L$ and $R$ respectively. Then, it counts the butterfly degree for all vertices and checks whether there exists two vertices $v_l \in V_L$ and $v_r \in V_R$ such that $\chi(v_l)\ge b$ and $\chi(v_r)\ge b$ hold.




Algorithm~\ref{algo:bccs} presents the details of finding $G_0$. For query vertices $Q$, first we pick out all vertices with the same labels with query vertices (line \ref{line:a1}). Each vertex set in $V_L$ and $V_R$ constructs the subgraph and we run the $k$-core algorithm respectively, find the connected component graph $L$ and $R$ containing query vertices $q_l$ and $q_r$ (lines \ref{line:a2}-\ref{line:a3}). Next we construct a bipartite graph $B$ to find cross-group butterfly structures in the community. $V_B$ consists of the vertex set $V_L$ and $V_R$, $E_B$ are cross-group edges (line \ref{line:a4}). Then, we compute the number of butterflies for each vertex in $B$ using Algorithm \ref{algo:count} (line \ref{line:a5}), which is presented in detail in the next paragraph.
Algorithm \ref{algo:count} returns the butterfly degree of all the vertices, maintaining two values $max_l$ and $max_r$ to record the maximum butterfly degree on each side. Then we check if there exists at least one vertex whose butterfly degree is no less than $b$ in each side, i.e., $max_l \geq b$ and $max_r \geq b$, otherwise return $\emptyset$ (lines \ref{line:a8}-\ref{line:a9}). Finally, we merge three subgraph parts to form $G_0$ (line \ref{line:a10}).

Next, we describe the details of the butterfly counting algorithm. $N_{v}^{2} = \{ w\ |\ (w, u) \in E \wedge w \neq v, \forall u \in N(v)\}$ is the set of vertices that are exactly in distance $2$ from $v$, i.e., neighbors of the neighbors of $v$ (excluding $v$ itself). To calculate the butterfly degree of each vertex, we take a vertex $v \in L$ as an example (the same for $v \in R$), each butterfly it participates in has one other vertex $w \in L (w \neq v)$ and two vertices $\{u, x\} \in R$. By definition, $w \in N_{v}^{2}$. In order to find the number of $u,x$ pairs that $v$ and $w$ form a butterfly, we compute the intersection of the neighbor sets of $v$ and $w$. 
We use $|\commonNei{v}{w}|$ to denote the number of common neighbors of $v$ and $w$, the number of the intersection pairs is $|\commonNei{v}{w}| \choose 2$. The butterfly degree equation for $v$ is as follows: $\chi(v) = \sum\nolimits_{u \in N(v)} {\sum\nolimits_{w \in N(u)\backslash v} {|\commonNei{v}{w}| \choose 2} } 
 = \sum\nolimits_{w \in N_v^2} {|\commonNei{v}{w}| \choose 2}$.
Instead of performing a set intersection at each step, we count and store the number of paths from a vertex $v \in L$ to each of its distance-$2$ neighbor $w \in L$ by using a hash map $P$ (lines \ref{line:a1}-\ref{line:a7}) \citep{sanei2018butterfly}. 

\begin{algorithm}[t]
	\small
	\caption{Butterfly Counting~\citep{sanei2018butterfly, wang2019vertex}} \label{algo:count}
	\begin{algorithmic}[1]
		\REQUIRE  $B=(V_B, E_B)$.
		\ENSURE $\chi(v)$ for all vertices $v \in V_B$.
		\FOR{$v \in V_B$}
		\STATE $P \gets hashmap$; // initialized with zero
		\FOR{$u \in N(v)$}
		\FOR{$w \in N(u)$}
		\STATE $P[w] \gets P[w] + 1$; //the number of 2-hop paths
		\ENDFOR
		\ENDFOR
		\FOR{$w \in P$}
		\STATE $\chi(v) \gets \chi(v) + {P[w] \choose 2}$;
		\ENDFOR
		\ENDFOR
		\RETURN $\{\chi(v) | \forall v \in V_B\}$;
	\end{algorithmic}
\end{algorithm}

\subsubsection{$(k_1, k_2, b)$-Butterfly-Core Maintenance}
Algorithm~\ref{algo:maintenance} describes the procedure for maintaining $G$ as a $(k_1, k_2, b)$-BCC after the deletion of vertices $S$ from $G$. In Algorithm~\ref{algo:Basic}, $S = {u^{*}}$ (line \ref{line:a8}). Generally speaking, after removing vertices $S$ and their incident edges from $G$, $G$ may not be a $(k_1, k_2, b)$-BCC any more, or $Q$ may be disconnected. Thus, Algorithm~\ref{algo:maintenance} iteratively deletes vertices having degree less than $k_1$ ($k_2$), until $G$ becomes a connected $(k_1, k_2, b)$-BCC containing $Q$. It firstly splits the vertex set $S$ to two parts by their labels (line \ref{line:a1}). Single vertex also works here, we only need to run Algorithm~\ref{algo:maintenance} on the corresponding side. Then run core maintenance algorithm on $L$ and $R$ respectively to maintain $L$ ($R$) as a $k_1$-core ($k_2$-core) (lines \ref{line:a2}-\ref{line:a3}). Next, we count butterfly degree again on the updated $B$ with Algorithm~\ref{algo:count} to check if it meets the butterfly constraint of the BCC model (line \ref{line:a4}). Finally, Algorithm~\ref{algo:maintenance} produces a BCC (line \ref{line:a5}).






\subsection{Approximation and Complexity Analysis}

We first analyze the approximation of Algorithm~\ref{algo:Basic}. 

\begin{theorem}\label{thm:Approximation}
Algorithm~\ref{algo:Basic} achieves $2$-approximation to an optimal solution $H^*$ of the BCC-problem, that is, the obtained $\{k_1, k_2, b\}$-BCC $\Result$ has $diam(\Result) \leq 2diam(H^*)$.
\end{theorem}
\vspace{-2mm}
\begin{proof}
First We have $diam(\Result)$ $=$ $\max_{u \in \Result, v \in \Result} $ $dist_\Result(u, v)$, and $dist_\Result(\Result, Q)$ $=$ $\max_{u \in \Result, v \in Q} dist_\Result(u, v)$, because of $Q \subseteq \Result$ then $dist_\Result(\Result, Q) \leq diam(\Result)$ holds. Suppose longest shortest path in $\Result$ is from vertex $s$ to $t$, i.e., $diam(\Result)= dist_\Result(s, t)$. For $\forall q \in Q$, $diam(\Result) = dist_\Result(s, t) \leq dist_\Result(s, q) + dist_\Result(q, t) \leq 2dist_\Result(\Result, Q)$. 

Next, we prove $dist_\Result(\Result, Q) \leq dist_{H^*}(H^*, Q)$ motivated by \citep{huang2015approximate}. 
Algorithm~\ref{algo:Basic} outputs a sequence of intermediate graphs $\{G_0,...,G_{l-1}\}$, 
which are BCC containing query vertices $Q$.  
$\Result$ is the one with the smallest query distance, i.e., $dist_\Result(\Result, Q) \leq dist_{G_i}(G_i, Q), \forall i \in \{0,...,l-1\}$. 
We consider two cases. 
(1) $H^* \nsubseteq G_{l-1}$. Suppose the first deleted vertex $v^* \in H^*$ happens in graph $G_i$, i.e., $H^* \subseteq G_i$, where $0 \leq i \leq l - 2$.
The vertex $v^*$ must be deleted because of the distance constraint but not the butterfly-core structure maintenance. 
Thus, $dist_{G_i}(G_i, Q) = dist_{G_i}(v^*, Q) = dist_{G_i}(H^*, Q)$. 
As a result, we have $dist_\Result(\Result, Q) \leq dist_{G_i}(G_i, Q) = dist_{G_i}(H^*, Q) \leq dist_{H^*}(H^*, Q)$, where the first inequality holds from $\Result$ has the smallest query distance, and the second inequality holds for that $H^*$ is a subgraph of $G_i$. 
(2) $H^* \subseteq G_{l-1}$. 
We prove $dist_{G_{l-1}}(G_{l-1}, Q) \leq dist_{G_{l-1}}(H^*, Q)$ by contradiction. 
This follows from the fact if $dist_{G_{l-1}}(G_{l-1}, Q) > dist_{G_{l-1}}(H^*, Q)$, 
then $G_{l-1}$ will not be the last feasible BCC.
There must exist a vertex $u^* \in G_{l-1} \backslash H^*$ with the largest query distance $dist_{G_{l-1}}(u^*, Q)$ so that $dist_{G_{l-1}}(u^*, Q) = dist_{G_{l-1}}(G_{l-1}, Q) > dist_{G_{l-1}}$ $(H^*, Q)$. 
In the next iteration, Algorithm~\ref{algo:Basic} will delete $u^*$ from $G_{l-1}$, and maintain the butterfly-core structure of $G_{l-1}$. As $H^*$ is a BCC, Algorithm~\ref{algo:Basic} can find a feasible BCC $G_l$ s.t. $H^* \subseteq G_l$, which contradicts that $G_{l-1}$ is the last feasible BCC. Overall, $dist_\Result(\Result, Q) \leq dist_{G_{l-1}}(G_{l-1}, Q) \leq dist_{G_{l-1}}(H^*, Q) \leq dist_{H^*}(H^*, Q)$.

From above we have proved that $diam(\Result) \leq 2dist_\Result(\Result, Q)$ and $dist_\Result(\Result, Q)$ $ \leq dist_{H^*}(H^*, Q)$. 
We have $diam(\Result) \leq 2dist_\Result(\Result, Q) \leq 2dist_{H^*}(H^*, Q) $ $\leq 2diam(H^*)$. 
\end{proof}
\vspace{-2mm}

To clarify, Algorithm~\ref{algo:Basic} returns a BCC community $\Result$ that has the minimum query distance $\Result = \mathop{\arg\min}_{G' \in \{G_0, ..., G_{l-1}\}} dist_{G'}(G', Q)$ (line~\ref{line:a10}), rather than the BCC community $O'$ with the smallest diameter, i.e, $O'= \mathop{\arg\min}_{G' \in \{G_0, ..., G_{l-1}\}} diam_{G'}(G', Q)$. This can speedup the efficiency by avoiding expensive diameter calculation. It is intuitive that the BCC community $O'$ also inherit 2-approximation  of the optimal answer, i.e., $diam(O')\leq diam(\Result) \leq 2diam(H^*)$.



\spara{Complexity analysis}. We analyze the time and space complexity of Algorithm~\ref{algo:Basic}. 
Let $t$ be the number of iterations and $t\leq\Nodenum$. We assume that $|V|-1 \leq |E|$, w.l.o.g., considering that $G$ is a connected graph.




\begin{algorithm}[t]
	\small
	\caption{Butterfly Core Maintenance$(G, S)$} \label{algo:maintenance}
	\begin{algorithmic}[1]
		\REQUIRE $G = L \cup B \cup R$, vertex set $S$ to be removed, three integers $\{k_1, k_2, b\}$.
		\ENSURE A $(k_1, k_2, b)$-BCC graph.
		\STATE Split $S$ into $S_L$ and $S_R$ according to their labels;
		\STATE Remove vertices $S_L$ from $L$ and maintain $L$ as $k_1$-core, update $B$;
		\STATE Remove vertices $S_R$ from $R$ and maintain $R$ as $k_2$-core, update $B$;
		\STATE Run butterfly Counting on $B$ and check there exists one vertex on each side with butterfly degree larger than $b$;
		\RETURN $G$;
	\end{algorithmic}
\end{algorithm}

\begin{theorem}
Algorithm~\ref{algo:Basic} takes $O(t (\sum\nolimits_{u \in V} {d_u^2} + \Edgenum))$ time and $O(\Edgenum)$ space. 
\end{theorem}

\vspace{-2mm}
\begin{proof}
Let $\Nodenum$ and $\Edgenum$ denote vertices and edges number respectively, $d_u$ is the vertex degree in the bipartite graph $B$. First, we consider the time complexity of finding $G_0$ in Algorithm \ref{algo:bccs} is $O(\sum\nolimits_{u \in B} {d_u^2} + \Edgenum)$ which runs $k$-core computation once in $O(\Edgenum)$ and calls Algorithm~\ref{algo:count} once in $O(\sum\nolimits_{u \in B} {d_u^2})$\citep{sanei2018butterfly}. Next, we analyze the time complexity of shrinking the community diameter. First, the computation of shortest distances by a BFS traversal starting from each query vertex $q \in Q$ takes $O(t|Q|\Edgenum)$ time for $t$ iterations, here $|Q| = 2$ then $O(t|Q|\Edgenum)$ could be eliminated to $O(t\Edgenum)$. Second, the time complexity of maintenance algorithm~\ref{algo:maintenance} is $O(t\sum\nolimits_{u \in B} {d_u^2} + \Edgenum)$ in $t$ iterations, i.e., $t$ times butterfly counting but note that $t$ times $k$-core maintain algorithm in total is $O(\Edgenum)$. The number of removed edges is no less than ($min\{k_1, k_2\} - 1$), thus the total number of iterations is $t \leq O(min\{\Nodenum - max\{k_1, k_2\}, \Edgenum/min\{k_1, k_2\}\})$ which is $O(\Nodenum)$ since $\Nodenum < \Edgenum$. As a result, the time complexity of Algorithm~\ref{algo:Basic} is $O(t ( \sum\nolimits_{u \in V} {d_u^2} + \Edgenum))$. Next, we analyze the space complexity of Algorithm \ref{algo:bccs}. It takes $O(\Nodenum + \Edgenum)$ space  to store graph $G$ and $O(\Nodenum)$ space to keep the vertex information including the coreness, butterfly degree, and query distance. Overall, Algorithm \ref{algo:bccs} takes $O(\Nodenum + \Edgenum) = O(\Edgenum)$ space, due to  a connected graph $G$ with $|V|-1 \leq |E|$.
\end{proof}
\vspace{-2mm}
\section{L$^2$P-BCC: Leader Pair based Local BCC Search}\label{sec:fast}
\begin{algorithm}[t]
\small
\caption{Fast Query Distance Computation$(G_i, q, D_i)$} \label{algo:sssp}
\begin{algorithmic}[1]
\REQUIRE A graph  $G_i$, query vertex $q$, a set of removal vertices $D_i$.
\ENSURE The updated distance $dist(v, q)$ for all vertices $v$.
\STATE Remove all vertices $D_i$ and their incident edges from $G_i$;
\STATE $d_{min} \gets min_{v \in D_i} dist_{G_i}(v, q)$;
\STATE $S_s = \{v\in V(G_i)\setminus D_i\ |\ dist_{G_i}(v, q) = d_{min} \}$;
\STATE $S_u = \{v\in V(G_i)\setminus D_i\ |\ dist_{G_i}(v, q) > d_{min} \}$;
\STATE Apply the BFS traverse on $G_i$ starting from $S_s$ to update the query distance of vertices in $S_u$;
\RETURN all updated distance $dist(u, q)$ for $u \in S_u$; \label{line:a22}
\end{algorithmic}
\end{algorithm}


Based on our greedy algorithmic framework in Algorithm~\ref{algo:Basic},  we propose three methods for fast BCC search in this section. 
The first method is the fast computation of query distance, which only updates a partial of vertices with new query distances in Section \ref{sub-sec:sp}. The second method fast identifies a pair of leader vertices, which both have the butterfly degrees no less than $b_p$ 
in Section \ref{sub-sec:lp}. The leader pair tends to have large butterfly degrees even after the phase of graph removal, which can 
save lots of computations in butterfly counting.
The third method of local BCC search is presented in Section \ref{sub-sec:lc}, which finds a small candidate graph for bulk removal refinement, instead of starting from whole graph $G$.

\vspace{-1mm}
\subsection{Fast Query Distance Computation}\label{sub-sec:sp}

Here, we present a fast algorithm to compute the query distance for vertices in $G$. In line $4$ of Algorithm~\ref{algo:Basic}, it needs to compute the query distance for all vertices, which invokes expensive computation costs. However, we observe that a majority of vertices keep the query distance unchanged after each phase of graph removal. This suggests that a partial update of query distances may ensure the updating exactness and speed up the efficiency. 

The key idea is to identify the vertices whose query distances need to update. Given a set of vertices $D_i$ deleted in graph $G_i$, let denote the distance $d_{min}= \min_{v\in D_i} dist_{G_i}(v, q)$. Let the vertex set $S_u = \{v\in V(G_i)\setminus D_i\ |\ dist_{G_i}(v, q) > d_{min} \}$. We have two importantly useful observations as follows. 

\squishlisttight

\item For each vertex $v\in S_u$, we need to recompute the query distance $dist_{G_{i+1}}(v, q)$. It is due to the fact that the vertex $u$ with $dist_{G_{i}}(u, q) <= d_{min}$ keeps the query distance unchanged in $G_{i+1}$. This is because no vertices that are along with the shortest paths between $u$ and $q$, are deleted in graph $G_{i}$. 

\item For vertex $v\in S_u$, $dist_{G_{i+1}}(v, q) \geq dist_{G_{i}}(v, q)$ always holds, due to $G_{i+1} \subseteq G_{i}$. Thus, instead of traversing from query vertex $q$, we update the query distance  $dist_{G_{i+1}}(v, q)$ by traversing from $S_s$, where $S_s = \{v\in V(G_i)\setminus D_i\ |\ dist_{G_i}(v, q) = d_{min} \}$.
\end{list}

We present the method of fast query distance computation in Algorithm~\ref{algo:sssp}. First, we remove a set of vertices $D_i$ from graph $G_i$ (line \ref{line:a1}). We then calculate the minimum distance $d_{min}$, which is the minimum length of the shortest path from $D_i$ to $q$ (line \ref{line:a2}). We select vertices $S_s$ whose shortest path equals to $d_{min}$ as the BFS starting points (line \ref{line:a3}). Let a set of vertices to be updated as $S_u$ whose shortest path is larger than $d_{min}$ (line \ref{line:a4}). Then, we run the BFS algorithm starting from vertices in $S_s$, we treat $S_u$ as unvisited and all the other vertices as visited (line \ref{line:a5}). The algorithm terminates until $S_u$ are visited or the BFS queue becomes empty. Finally, we return the shortest path to $q$ for all vertices (line \ref{line:a6}). Note that in each iteration of Algorithm~\ref{algo:Basic} it always keeps one query vertex's distance to other vertices unchanged, only needs to update the distances of another query vertex, because $S_u = \emptyset$ always holds for one of the query vertices.

\begin{table}[t]
	\centering
	\small
	\caption{The shortest distances of queries to other vertices. The symbol ``$-$'' represents the vertex set unchanged. The vertices with changed distance are depicted in bold, i.e., $u_4$ and $u_7$.}
	\vspace{-0.2cm}
	\label{query_distance}
	\begin{tabular}{ l|rrrrr }
		\hline
		{\bf query} & $1$ & $2$ & $3$ & $4$  \\
		\hline
		$q_l$ & $\{v_1, v_2, v_3\}$ & $\{u_2, u_3, u_5, u_6\}$ & $\{q_r, u_1, u_4, u_7\}$ & $\{u_9\}$\\
		$q_r$ & $\{u_1, u_2, u_3, u_9\}$ & $\{v_1, v_3, u_4, u_5, u_7\}$ & $\{q_l, v_2, u_6\}$ & $\emptyset$ \\ \hline
		\multicolumn{5}{c}{\bf after the deletion of $u_9$} \\ \hline
		$q_l$ & $-$ & $-$ & $-$ & $\emptyset$\\
		$q_r$ & $\{u_1, u_2, u_3\}$ & $\{v_1, v_3, u_5\}$ & $\{q_l, v_2, u_6, \boldsymbol{u_4, u_7}\}$ & $\emptyset$ \\ \hline
	\end{tabular}
	\vspace{-0.6cm}
\end{table}

\begin{figure}[t]
	\vspace*{0.4cm}
	\centering 
	\includegraphics[width=0.8\linewidth]{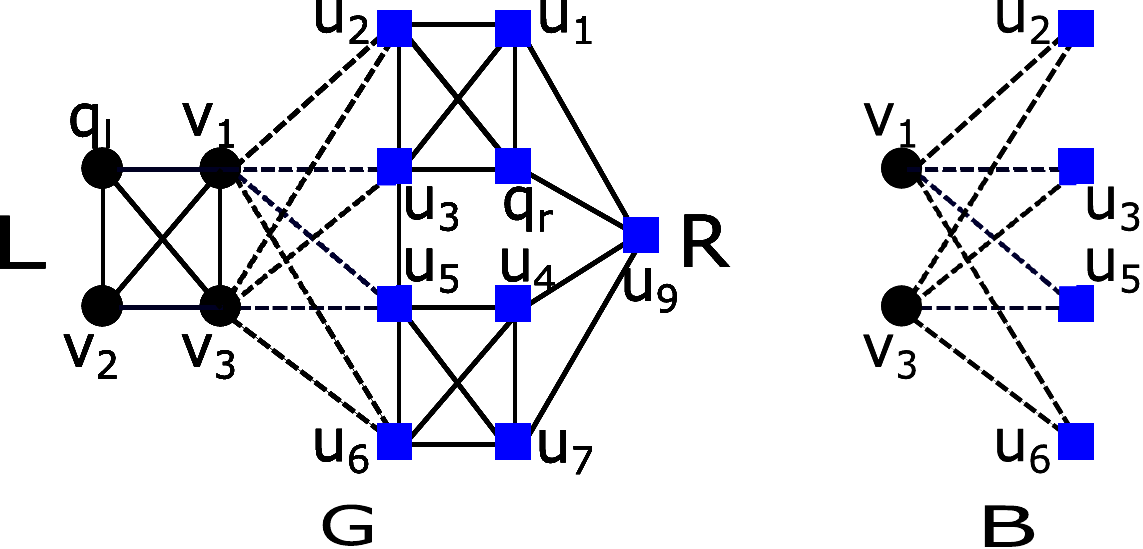}
	\vspace*{-0.3cm}
	\caption{An example of labeled graph $G$ and its bipartite subgraph $B$.}
	\vspace*{-0.4cm}
	\label{fig.fast}
\end{figure}

\begin{example}
Consider the graph $G$ in Figure~\ref{fig.fast} as $G_i$ and $Q = \{q_l, q_r\}$. From Table~\ref{query_distance} we know the vertex $u_9$ has the maximum distance to $Q$, i.e., $dist_{G}(u_9, Q) = 4$ (line~\ref{line:a5} in Algorithm~\ref{algo:Basic}). Thus, the removal vertex set is $D_i = \{u_9\}$. Now, we apply Algorithm~\ref{algo:sssp} on $G$ to update the query distance. (1) For $q_l$, the vertex $u_9$ is the farthest vertex to $q_l$, thus $S_u = \emptyset$ (line~\ref{line:a4}), which indicates no vertices' distances to $q_l$ need to update. (2) For $q_r$, $D_i = \{u_9\}$ and $d_{min} = min_{v \in \{u_9\}} dist_G (v, q_r) = dist_G (u_9, q_r) = 1$ (line~\ref{line:a2}). The vertex set to update is $S_u = \{v\in V(G)\setminus \{u_9\}\ |\ dist_{G}(v, q_r) > 1 \} = \{v\in V(G)\setminus \{u_9\}\ |\ dist_{G}(v, q_r) = 2\} \cup \{v\in V(G)\setminus \{u_9\}\ |\ dist_{G}(v, q_r) = 3 \} = \{v_1, v_3, u_4, u_5, u_7\} \cup \{ q_l, v_2, u_6\} = \{q_l, v_1, v_2, v_3, u_4, u_5, u_6, u_7\}$ (line~\ref{line:a4}). Then, we apply BFS search starting from the vertex set $S_s = \{v \in V(G)\setminus\{u_9\}\ |\ dist_{G}(v, q_r) = 1\} = \{u_1, u_2, u_3\}$ (line~\ref{line:a3}), and update the shortest distance for vertices in $S_u$ (line~\ref{line:a5}). The updated distances are also shown in Table~\ref{query_distance}. 
\end{example}

\subsection{Leader Pair Identification}\label{sub-sec:lp}
Next, we present efficient algorithms for leader pair identification, to improve the efficiency of the butterfly counting part. 

\spara{Leader pair identification}. The butterfly counting in Algorithm~\ref{algo:count} needs to be invoked once a graph $G_i$ is updated in Algorithm~\ref{algo:Basic}. This may lead to a large number of butterfly counting, which is very time costly. Recall that the definition of BCC requires a pair of vertices $\{v_l\in L, v_r \in R\}$ such that $\chi(v_l) \geq b$ and $\chi(v_r) \geq b$. This motivates us to find a good pair of leader vertices, whose butterfly degrees are large enough w.r.t. $b_p$ in two groups $L$ and $R$, even after a number of graph removal iterations. Thus, it can avoid finding a new leader pair and save time cost. In the following, we show two key observations to find a good leader pair $(v_l, v_r)$ where $v_l\in L$ and $v_r \in R$.

\begin{observation}
The leader pair $P$ should have large butterfly degrees $\chi(v_l)$ and $\chi(v_r)$, which do not easily violate the constraint of cross-over interactions.  
\end{observation}

\begin{observation}
The leader pair $P$ should have small query distances $dist(v_l, Q)$ and $dist(v_r, Q)$, which are  close to query vertices $Q$ and not easily deleted by graph removal.
\end{observation}


We present the algorithm of leader pair identification in Algorithm~\ref{algo:LeaderPair}. Here, we use the graph $G$ to represent a graph $L/R$ and find a leader vertex $v_l/v_r$, $\lp$ denotes to search leaders within $\lp$-hops neighbors of the query vertex $q$. We first initiate $p$ as the query vertex $q$ since it is the closest with distance $0$ (line~\ref{line:a1}). This is especially effective when the input query vertex is leader biased who contains the largest butterfly degree. If the degree number is large enough, i.e., greater than $b_{max}/2$, we return $p$ as the leader vertex (lines~\ref{line:a2}-\ref{line:a5}); Otherwise, we find the leader vertex $p$ in the following manner. We first increase $d$ from $1$ to $\lp$ (line \ref{line:a14}), and decrease $b_p$ in $\{b_{max}/2, b_{max}/4, ..., b\}$ (lines~\ref{line:a7}-\ref{line:a15}). We get the set of vertices $S$ whose distance to the query is $d$ (line \ref{line:a10}). Then, we identify one vertex $s\in S$ with $\chi(s)\geq b_p$ and return as the leader vertex otherwise we increase $d$ by $1$ to search the next hop (lines \ref{line:a9}-\ref{line:a14}). Note that we return an initial $p$ if no better answer is identified (line \ref{line:a16}).


\begin{algorithm}[t]
	\small
	\caption{Leader Pair Identification $(G, q, \lp, b)$} \label{algo:LeaderPair}
	\begin{algorithmic}[1]
		\REQUIRE A graph $G = L$ (or $R$), a query vertex $q = q_l$ (or $q_r$), $\lp$, $b$.
		\ENSURE A lead vertex $p=v_l$ (or $v_r$).
		\STATE $p \gets q$; //Initiate as query vertex.
		\STATE $b_{max} \gets max_{v \in G} \chi(v)$;
		\STATE $b_p \gets b_{max}/2$;
		
		\IF{$\chi(p) > b_p$}
		\RETURN $p$; //Query vertex has a large enough butterfly degree.
		\ELSE
		
		\WHILE{$b_p \geq b$}
		\STATE $d \gets 1$; //Search from the $1$-hop neighbors of $q$. 
		\WHILE{$d \leq \lp$}
		\STATE $S \gets \{u\ |\ dist_{G} (u, q) = d, u \in G\}$;
		\IF{$\exists s \in S$ that $\chi(s) \geq b_p$} \label{line:a11}
		\RETURN  $s$; \label{line:a12}
		\ELSE \label{line:a13}
		\STATE $d \gets d + 1$; \label{line:a14} //Increase the search distance $d$.
		\ENDIF
		\ENDWHILE
		\STATE $b_p \gets b_p/2$; \label{line:a15}
		\ENDWHILE 
		\ENDIF
		\RETURN $p$; \label{line:a16}
	\end{algorithmic}
\end{algorithm}

	\begin{example}\label{example:LeaderPair}
		We apply Algorithm~\ref{algo:LeaderPair} on $G$ in Figure~\ref{fig.fast} for $Q = \{q_l, q_r\}$ and select $\lp = 3$. In the graph $G$, the non-zero butterfly degree of vertices are $\chi(v_1) = \chi(v_3) = 6$ and $\chi(u_2) = \chi(u_3) = \chi(u_5) = \chi(u_6) = 3$. (1) For the subgraph $L$, first it initializes $p$ as the query vertex $q_l$ (line~\ref{line:a1}). Since $b_{max} = 6$ and $b_p = b_{max}/2 = 3$ (lines~\ref{line:a2}-\ref{line:a3}), where $\chi(p) = 0$ which is less than $b_p = 3$ then it goes to line~\ref{line:a6}. Then, it starts from $d = 1$ (line~\ref{line:a8}) to search $q_l$'s $1$-hop neighbors, i.e., $S = \{v_1, v_2, v_3\}$ (line~\ref{line:a10}), finally finds there exists vertex $s = v_1$ such that $\chi(s) \geq b_p = 3$ and returns $v_1$ as the leader vertex (lines~\ref{line:a11}-\ref{line:a12}). (2) For the subgraph $R$, it follows a similar process where $b_{max} = 3$ and $b_p = b_{max}/2 = 1.5$, returns $u_2$ as the leader vertex. Finally, we obtain $\{v_1, u_2\}$ as the leader pair.
	\end{example}




\spara{Butterfly degree update for leader pair}. Here, we consider how to efficiently update the butterfly degrees of leader pair vertices $\chi(v_l)$ and $\chi(v_r)$ after graph removal in Algorithm~\ref{algo:Basic}. 

The algorithm of updating the butterfly degrees of the leader pair is outlined in Algorithm~\ref{algo:UpdatePair}. First, we check the labels of $p$ and $v$ (line \ref{line:a1}). If $\ell(p) = \ell(v)$, we find the common neighbors $\commonNei{v}{p}$ shared by $p$ and $v$, its number denoted as $\cn$, then the number of butterflies containing $p$ and $v$ is ${\cn \choose 2}$. Then, $\chi(p)$ decreases by ${\cn \choose 2}$ (lines \ref{line:a2}-\ref{line:a3}). If $\ell(p) \neq \ell(v)$, firstly there exists no butterflies if $p$ and $v$ do not connect (line \ref{line:a5}). We enumerate each vertex $u$, i.e., $v$'s neighbors, and check their common neighbors with $p$, i.e., $\commonNei{u}{p}$. Note that $\bcount$ keeps the number of butterflies involving $v$, so we directly update $\chi(p)$ by decreasing $\bcount$ (lines \ref{line:a6}-\ref{line:a8}).


	\begin{example}
We apply Algorithm~\ref{algo:UpdatePair} on $G$ in Figure~\ref{fig.fast} to update the butterfly degree of leader pair $\{v_1, u_2\}$. First, the deletion of vertex $u_9$ has no influence on the butterfly degree. Next vertex $u^*$ to delete is selected from $\{v_2, u_1, u_4, u_6, u_7\}$, which has the maximum query distance $dist_{G}(u^*, Q) = 3$. To illustrate, we assume to delete $u_6$. (1) For $u_2$, $\chi(u_2) = 3$ before the deletion. Since $\ell(u_2) = \ell(u_6)$ (line~\ref{line:a1}), their common neighbors $\commonNei{u_2}{u_6} = \{v_1, v_3\}$ and $\cn = |\{v_1, v_3\}| = 2$ (line~\ref{line:a2}). The updated butterfly degree is $\chi(u_2) = 3 - {\cn \choose 2} = 3 - 1 = 2$ (line~\ref{line:a3}). (2) For $v_1$, $\chi(v_1) = 6$ before the deletion. Since $\ell(v_1) \neq \ell(u_6)$ and $u_6 \in N(v_1)$ (lines~\ref{line:a4}-\ref{line:a5}), we enumerate $u_6$'s neighbors except $v_1$, i.e., $\{v_3\}$~(line~\ref{line:a6}). Since $\bcount = \sum_{u \in \{v_3\}} (|\commonNei{u}{v_1}| - 1) = |\commonNei{v_3}{v_1}| - 1 = |\{u_2, u_3, u_5, u_6\}| - 1 = 3$~(line~\ref{line:a7}), then the updated butterfly degree is $\chi(v_1) = 6 - 3 = 3$~(line~\ref{line:a8}).
	\end{example}


\begin{algorithm}[t]
	\small
	\caption{Butterfly Degree Update for Leader Pair $(B, p, v)$} \label{algo:UpdatePair}
	\begin{algorithmic}[1]
		\REQUIRE A graph $B = (V_B, E_B)$, a leader vertex $p$, a deletion vertex $v$.
		\ENSURE Butterfly degree $\chi(p)$.
		\IF{$\ell(p) = \ell(v)$}
		\STATE $\cn \gets |\commonNei{v}{p}|$; //The number of common neighbors of $p$, $v$.
		\STATE $\chi(p) \gets \chi(p) - {\cn \choose 2}$;
		\ELSE
		\IF{$v \in N(p)$}
		\FOR{$u \in N(v)$ \AND  $u \neq p$}
		\STATE $\bcount \gets \bcount + |\commonNei{u}{p}| - 1$; //Add the count of removed butterflies. 
		\ENDFOR
		\STATE $\chi(p) \gets \chi(p) - \bcount$;
		\ENDIF
		\ENDIF
		\RETURN $\chi(p)$;
	\end{algorithmic}
\end{algorithm}

\spara{Complexity analysis}. Next, we analyze the time and space complexity of leader pair 
identification and update in 
Algorithms~\ref{algo:LeaderPair} and~\ref{algo:UpdatePair}. First, Algorithm~\ref{algo:LeaderPair} takes $O(\Edgenum\log{(b_{max}-b)})$ time in $O(\Edgenum)$ space, which identifies the leader pair using a binary search of approximate butterfly degree $b_p\in[b, b_{max}]$ within the query vertex's neighborhood. 
Next, we analyze the leader pair  update in Algorithm~\ref{algo:UpdatePair}.

\begin{theorem}
One run of butterfly degree update in Algorithm~\ref{algo:UpdatePair} takes $O(d_u^2)$ time and $O(\Edgenum)$ space, where $d_u = max_{v \in B} d(v)$
\end{theorem}

\vspace{-2mm}
\begin{proof}
The degree of leader vertex and delete vertex are $d_p$ and $d_v$, the time complexity of Algorithm~\ref{algo:UpdatePair} is $O(min(d_p, d_v))$ if $\ell(p) = \ell(v)$; $O(d_p d_v)$ if $\ell(p) \neq \ell(v)$. Assume that the maximum degree in bipartite graph $B$ is $d_u$. Thus, the total complexity of Algorithm~\ref{algo:UpdatePair} is $O(d_u^2)$.  In addition, we analyze the space complexity. Algorithms~\ref{algo:UpdatePair}  take $O(\Nodenum + \Edgenum) = O(\Edgenum)$ space to store the vertices and their incident edges in the graph.
\end{proof}
\vspace{-2mm}

In summary, a successful leader pair identification by Algorithm~\ref{algo:LeaderPair} can significantly reduce the times of calling the butterfly counting in Algorithm~\ref{algo:count}. In addition, it only needs to update the butterfly degree of leader vertices using Algorithm~\ref{algo:UpdatePair} but not the entire vertex set in BCC candidate graph, which is also an improvement of butterfly counting. This butterfly computing strategy for leader pair identification and update is very efficient for BCC search, as validated in Exp-5 in Section~\ref{sec:exp}.

\subsection{Index-based Local Exploration}\label{sub-sec:lc}
In this section, we develop a query processing algorithm which efficiently detects a small 
 subgraph candidate around query vertices $Q$, which tends to be densely connected both in its own label subgraph and bipartite graph.

First, we construct the BCindex for all vertices in $G$. The data structure of BCindex consists of two components: the coreness and butterfly number. For $k$-core index construction of each vertex, we apply the existing core decomposition \citep{batagelj2003m} on graph $G$ to construct its coreness. The $k$-core has a nested property, i.e., the $k_1$-core of graph $G$ must be a subgraph of $k_2$-core if $k_1 > k_2$. The offline $k$-core index could efficiently find $k$-core subgraphs from $G$, meanwhile reduce the size of bipartite graph $B$. Moreover, we keep the butterfly degree number index of each vertex on the bipartite graph with different labels using Algorithm~\ref{algo:count}.


Based on the obtained BCindex, we present our method of index-based local exploration, which is briefly presented in Algorithm~\ref{algo:locbcc}. The algorithm starts from the query vertices and finds the shortest path connecting two query vertices. A naive method of shortest path search is to find a path is the minimum number of edges, while this may produce a path along with the vertices of small corenesses and small butterfly degrees. To this end, we give a new definition of butterfly-core path weight as follows.

%

\begin{algorithm}[t]
	\small
	\caption{Index-based Local Exploration} \label{algo:locbcc}
	\begin{algorithmic}[1]
		\REQUIRE $G = L \cup B \cup R$, $Q = \{q_l, q_r\}$, three integers $\{k_1, k_2, b\}$.
		\ENSURE A connected $(k_1, k_2, b)$-BCC $R$ with a small diameter.
		\STATE Compute a shortest path $P$ connecting $Q$ using the butterfly-core path weight;
		\STATE $k_l \gets \min_{v \in V_L} \delta(v)$; $k_r \gets \min_{v \in V_R} \delta(v)$;
		\STATE Iteratively expand $P$ into graph $G_t = \{v \in L\ |\ \delta(v) \geq k_l \} \cup \{v \in R\ |\ \delta(v) \geq k_r \}$ by adding adjacent vertices $v$, until $|V(G_t)| > \eta$;
		\STATE Compute a connected $(k_1, k_2, b)$-BCC containing $Q$ of $G_t$ with the largest coreness on each side;
		\STATE Remove disqualified subgraphs on $G_t$ to identify the final BCC;
	\end{algorithmic}
\end{algorithm}
\vspace{-2mm}

\begin{definition}[Butterfly-Core Path Weight]
Given a path $P$ between two vertices $s$ and $t$ in $G$, the butterfly-core weight of path $P$ is defined as $\underline{dist}(s, t) = dist_G(s,t) + \gamma_1 (\delta_{max} - \min_{v \in P} \delta(v)) + \gamma_2 (\chi_{max} - \min_{v \in P}\chi(v))$, where $\delta(v)$ is the coreness of vertex $v$, $\chi(v)$ is the butterfly degree of $v$, $\delta_{max}$ and $\chi_{max}$ are the maximum coreness and the maximum butterfly degree in $G$.
\end{definition}
\vspace{-2mm}

The value of $(\delta_{max} - \min_{v \in P} \delta(v))$ and $(\chi_{max} - \min_{v \in P}\chi(v))$ respectively measures the shortfall in the coreness and butterfly degree of vertices in path $P$ w.r.t. the corresponding maximum value in $G$. Smaller the shortfall of a path, lower its distance. Here, $\gamma_1$ controls the extent to which a small vertex coreness is penalized, and $\gamma_2$ controls the penalized extent of a small butterfly degree. Using BCindex, for any vertex $v$ we can access the structural coreness $\delta(v)$ and the butterfly degree $\chi(v)$ in $O(1)$ time. 
Thus, the shortest path $P$ can be extracted based on the butterfly-core path weight definition.


We then expand the extracted path $P$ to a large graph $G_t$ as a candidate BCC by involving the local neighborhood of query vertices. We start from vertices in $P$, split the vertices by their labels into $V_L$ and $V_R$, obtain the minimum coreness of vertices in each side as $k_l = \min_{v \in V_L} \delta(v)$ and $k_r = \min_{v \in V_R} \delta(v)$ (line \ref{line:a2}). Due to the different density of two groups, we expand $P$ in different core values. We then expand the path by iteratively inserting adjacent vertices with coreness no less than $k_l$ and $k_r$ respectively, in a BFS manner into $G_t$, until the vertex size exceeds a threshold $\eta$, i.e., $|V(G_t)| > \eta$, where $\eta$ is empirically tuned. After that, for each vertex $v \in V (G_t)$, we add all its adjacent edges into $G_t$ (line \ref{line:a3}).  Since $G_t$ is a local expansion, the coreness of $L$, $R$ will be at most $k_l$ and $k_r$. Based on $G_t$, we extract the connected $(k_1, k_2, b)$-BCC containing $Q$. If input parameters $k_1$ and $k_2$ are not supplied, they are set automatically with the largest coreness on each side (line \ref{line:a4}). Then it iteratively removes the farthest vertices from this BCC. Moreover, since $dist_{G_i}(u, Q)$ is monotone non-decreasing with decreasing graphs, to reduce the iteration number of butterfly-core maintenance, we propose to delete a batch of vertices with the same farthest query distance, i.e., $S = \{u^* | dist_{G_i}(u^*, Q) \geq d_{max}\}$, which can further improve the search efficiency (line \ref{line:a5}). Although Algorithm~\ref{algo:locbcc} does not preserve $2$-approximation guarantee of optimal answers, it achieves the results of good quality fast in practice, as validated in our experiments.


\section{Handle Multi-labeled BCC Search}\label{sec:multiple}

In this section, we generalize the BCC model from $2$ group labels to multiple $m$ group labels where $m\geq2$, and formulate the multi-labeled BCC-problem. Then, we discuss how to extend our existing techniques to handle a BCC search query with multiple labels.

\stitle{Multi-labeled BCC search}. First, we extend the \emph{cross-group interaction} to  a new definition of \emph{cross-group connectivity}. 
According to the 4th constraint of Def.~\ref{def.bcc}, two groups labeled with $A_l$ and $A_r$ have cross-group interaction \emph{iff} $\exists v_l \in V_L$ and $\exists v_r\in V_R$ such that butterfly degree $\chi(v_l) \geq b$ and $\chi(v_r) \geq b$ in the induced subgraph of $H$ by $V_L\cup V_R$. We say that two labels $A_l$ and $A_r$ have cross-group interaction, denoted as $(A_l, A_r)$. 

\begin{definition}
[Cross-group Connectivity] Two labels $A_l$ and $A_r$ have cross-group connectivity if and only if there exists a \emph{cross-group path} $P=\{A_1, \ldots, A_j\}$ where $A_1=A_l$, $A_j=A_r$, and $(A_{i}, A_{i+1})$ has cross-group interaction for any $1\leq i <j$. 
\label{def.mconnectivity}
\end{definition}

To model group connection in a multi-labeled BCC, it requires that for any two labels $A_l$ and $A_r$, there exists either a direct \emph{cross-group interaction} or a \emph{cross-group path} between two group core structures, implying the strong connection between two different groups within a BCC community. Specifically, we extend the definition of the multi-labeled BCC (mBCC) model as follows. 


\begin{definition}[Multi-labeled Butterfly-Core Community]
Given a labeled graph $G$, an integer $\MultipleLabels \geq 2$, group core parameters $\{k_i: 1\leq i\leq \MultipleLabels\}$, a multi-labeled butterfly-core community (mBCC) $H \subseteq G$ satisfies the following conditions:

1. Multiple labels: there exist $\MultipleLabels$ different labels $A_1, A_2, ..., A_\MultipleLabels \in \mathcal{A}$ such that $\forall 1\leq i\leq m$, $V_i =\{v\in V(H): \ell(v)=A_i\}$ and $V_i\cap V_j=\emptyset$ for $i\neq j$, and $V_1\cup V_2\cup ...\cup V_\MultipleLabels = V_H$;


2. Core groups: the induced subgraph of $H$ by $V_i$ is $k_i$-core where $1\leq i \leq \MultipleLabels$; 

3. Cross-group connectivity: for $1\leq i, j\leq \MultipleLabels$, any two labels $A_i$ and $A_j$ have cross-group connectivity in $H$.


\label{def.mbcc}
\end{definition}

For $\MultipleLabels=2$, this definition is equivalent to our butterfly-core community model in Def.~\ref{def.bcc}. From the conditions (1) and (2) the mBCC has exactly $\MultipleLabels$ labeled groups and each group is a $k_i$-core. The condition (3) requires that these $\MultipleLabels$ groups could be cross-group connected by the bipartite interactions. To ensure the cross-group connectivity,  each group should have one leader vertex whose  butterfly degree has no less than  $b$. Note that the accumulated number of butterflies from multiple bipartite graphs does not all count into the butterfly degree. Specifically, we only count on the butterfly degree among two labeled bipartite graph. 

\stitle{mBCC-search problem}.  Based on the multi-labeled BCC community model, the corresponding mBCC-search problem can be similarly defined as Problem~\ref{def.bcc_problem} as follows. Consider a query $Q=\{q_1, \ldots, q_\MultipleLabels\}$ where each query vertex $q_i$ has a distinct label $A_i$, the group core parameters $\{k_i: 1\leq i\leq \MultipleLabels\}$, and the butterfly degree parameter $b$, the \emph{mBCC-search problem} is to find a connected mBCC containing $Q$ with the smallest diameter.

\begin{algorithm}[t]
	\small
	\caption{Multi-labeled BCC Search Framework $(G, Q)$} \label{algo:Multiple}
	\begin{algorithmic}[1]
		\REQUIRE $G = (V, E, \ell)$, multi-labeled query $Q=\{q_1, \ldots, q_\MultipleLabels\}$, group core parameters $\{k_i: 1\leq i\leq \MultipleLabels\}$, butterfly degree parameter $b$.
		\ENSURE A connected mBCC $\Result$ with a small diameter.
		\STATE Find a maximal connected subgraph of multi-labeled BCC containing $Q$ as $G_0$ by Def.~\ref{def.mbcc}; //using Alg.~\ref{algo:bccs} \label{line:a1}
		\STATE $l \gets 0$; \label{line:a2}
		\WHILE{$connect_{G_l} (Q) = \textbf{true}$} \label{line:a3}
		\STATE Delete vertex $u^*\in V(G_l)$ with the largest $dist_{G_l} (u^*, Q)$ from $G_l$; //using Alg.~\ref{algo:sssp} \label{line:a4}
		\STATE Maintain each labeled group as a $k_i$-core, $\forall i \in \{1, 2, ...,\MultipleLabels\}$;  \label{line:a5}
		\STATE Update the leader pairs and check the cross-group connectivity among $m$ labeled groups by Def.~\ref{def.mconnectivity}; //Alg.~\ref{algo:count}~\&~\ref{algo:maintenance}, optimized by Alg.~\ref{algo:LeaderPair}~\&~\ref{algo:UpdatePair} \label{line:a6}
		\STATE  $G_{l+1} \gets G_l$; $l \gets l + 1$; \label{line:a7}
		\ENDWHILE
		\STATE $\Result \gets \mathop{\arg\min}_{G' \in \{G_0, ..., G_{l-1}\}} dist_{G'}(G', Q)$; \label{line:a8}
	\end{algorithmic}
\end{algorithm}

\stitle{Algorithm}. We propose a multi-labeled BCC search framework in Algorithm~\ref{algo:Multiple}, which leverages our previous techniques of search framework in  Algorithm~\ref{algo:Basic}  and fast strategies in Section~\ref{sec:fast}. 
The algorithm first finds a maximal connected mBCC containing all query vertices $Q$ as $G_0$ (line~\ref{line:a1}) and then iteratively removes the farthest vertex $u^*$ from the graph (lines~\ref{line:a3}-\ref{line:a7}). The  mechanism is to maintain each labeled group as a $k_i$-core (line~\ref{line:a5}) and update the leader pairs (line~\ref{line:a6}). Based on the identified leader pairs and cross-group interactions, the algorithm can check the cross-group connectivity for $m$ labeled groups in $G_l$ by Def.~\ref{def.mconnectivity} as follows. We first construct a new graph $H_m$ with $m$ isolated vertices, where each vertex represents a labeled group. Then, we insert into $H_m$ an edge between two vertices if two labeled groups have a cross-group interaction. Finally, the cross-group connectivity of $G_l$ is equivalent to the graph connectivity of $H_m$. In addition, our fast query distance computation in Algorithm~\ref{algo:sssp} and local search strategies in Algorithms~\ref{algo:LeaderPair} and~\ref{algo:UpdatePair} can be also easily extended to improve the efficiency. 

\stitle{Complexity analysis}. We analyze the complexity of Algorithm~\ref{algo:Multiple}. At each iteration of Algorithm~\ref{algo:Multiple}, the shortest path computation takes $O(|Q|\cdot(|V|+|E|)) = O(m|E|)$ time, due to $|V| -1 \leq |E|$.  Moreover, it takes $O(|V|+|E|)$ time to find and maintain all $k_i$-cores for different $m$ labels. To identify the leader pairs, it runs  the butterfly counting over the whole graph once in $O(\sum\nolimits_{u \in V} {d_u^2})$ time. The extra cost of checking cross-group connectivity takes $O(m^2)\subseteq O(m|E|)$ time, due to the queries $Q\subseteq V$ and $|V| -1 \leq |E|$. Actually, the checking cross-group connectivity can be further optimized in $O(m)$ time using the union-find algorithm in any conceivable application~\citep{cormen2009introduction}. Thus, for a query $Q$ with $m$ different labeled vertices, Algorithm~\ref{algo:Multiple} takes $O(t(m|E|+\sum\nolimits_{u \in V} {d_u^2}))$ time for $t$ iterations and $O(|E|)$ space.




\section{Experiments}\label{sec:exp}
In this section, we conduct experiments to evaluate our proposed model and algorithms.


\begin{table}[t]
\centering
\small
\vspace{-0.3cm}
\caption{Network statistics $(\bf{K} = 10^3$ and $\bf{M} = 10^6)$.}
\vspace{-0.2cm}
\label{snap}
\begin{tabular}{ l|rrrrr }
\hline
{\bf Network} & $|V|$ & $|E|$ & $Labels$ & $k_{max}$ & $d_{max}$  \\
 \hline
\Quarter & 30\bf{K} & 508\bf{K} & 383  & 43 & 12\\
\Year & 41\bf{K} & 2\bf{M} & 346  & 189 & 13\\
Amazon  & 335\bf{K} & 926\bf{K} & 2  & 6 & 549\\
DBLP  & 317\bf{K} & 1\bf{M} & 2  & 113 & 342\\
Youtube & 1.1\bf{M} & 3\bf{M} & 2  & 51 & 28,754\\
LiveJournal & 4\bf{M} & 35\bf{M} & 2  & 360 &14,815\\
Orkut & 3.1\bf{M} & 117\bf{M} & 2  & 253 & 33,313\\ \hline
\end{tabular}
\vspace{-0.5cm}
\end{table}


\spara{Datasets}. We use seven real datasets as shown in Table~\ref{snap}. Two new real-world datasets of labeled graphs are collected from Baidu which is a high tech company in China. They are IT professional networks with the ground-truth communities of joint projects between two department teams, denoted as \emph{\Quarter} and \emph{\Year}. In both graphs, each vertex represents an employee and the label represents the working department. An edge exists between two employees if they have communication through the company's internal instant messaging software. \Quarter and \Year are generated based on data logs for three months and one whole year, respectively. In addition, we use five graphs with ground-truth communities from SNAP, namely Amazon, DBLP, Youtube, LiveJournal and Orkut, with randomly added synthetic vertex labels into them. Specifically, we split the vertices based on communities into two parts, assigned all vertices in each part with one label.  We also generated the query pairs by picking any two vertices with different labels. To add cross edges within communities, we randomly assigned vertices with 10\% cross edges to simulate the collaboration behaviors between two communities. Moreover, we added 10\% noise data of cross edges globally on the whole graph.

\spara{Compared methods.} We compare our three BCC search approaches with two community search competitors as follows. 

\squishlisttight
\item \CTC: 
finds a closest $k$-truss community containing a set of query vertices~\citep{huang2015approximate}.

\item \PSA: 
progressively finds a minimum $k$-core containing a set of query vertices~\citep{li2019efficient}.

\item \BD: our online BCC search in Algorithm \ref{algo:Basic}.

\item \BDQL: our \BD method equipped with two accelerate strategies, i.e., fast query distance computation in Algorithm~\ref{algo:sssp} and \underline{l}eader \underline{p}air identification in Algorithms~\ref{algo:LeaderPair} $\&$~\ref{algo:UpdatePair}.  

\item \LBCC: our \underline{l}ocal \underline{l}eader \underline{p}air BCC search described in Algorithm~\ref{algo:locbcc}, which is also equipped with Algorithm~\ref{algo:sssp} and Algorithms~\ref{algo:LeaderPair} $\&$~\ref{algo:UpdatePair}.

\end{list}








Note that all our methods use bulk deletion that removes a batch of vertices with the farthest distances from the graph.
We use the default parameter setting of \CTC \citep{huang2015approximate} and \PSA \citep{li2019efficient}. For \LBCC, we set $\gamma_1 = 0.5$ and $\gamma_2 = 0.5$. 

\spara{Evaluation metrics.}
To evaluate the efficiency, we report the running time in seconds and treat a query search as infinite if it exceeds $30$ minutes. To evaluate the community quality, we use the F1-score metric to measure the alignment between a discovered community $C$ and a ground-truth community $\hat{C}$. Here, F1 is defined as $F1(C, \hat{C}) = \frac{2·prec(C, \hat{C})·recall(C, \hat{C})}{prec(C, \hat{C}) + recall(C, \hat{C})}$ where $prec(C, \hat{C}) = \frac{|C \cap \hat{C}|}{|C|}$ and $recall(C, \hat{C}) = \frac{|C \cap \hat{C}|}{|\hat{C}|}$ .

\begin{figure}[t]
\centering
\includegraphics[width=\linewidth]{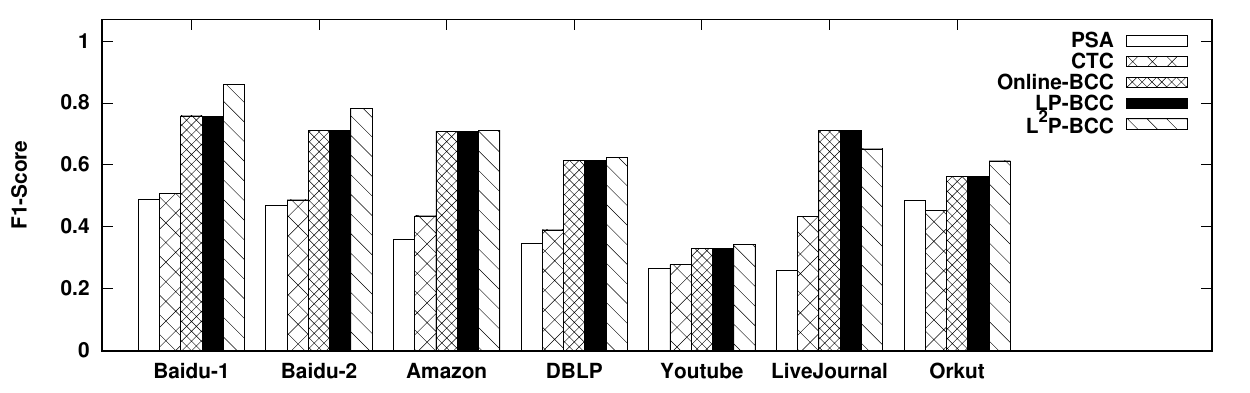}
\vspace{-9mm}
\caption{Quality evaluation on networks with ground-truth communities.}
\vspace{-2mm}
\label{fig.f1}
\end{figure}

\begin{figure}[t]
	\vspace{-2mm}
	\centering
	\includegraphics[width=\linewidth]{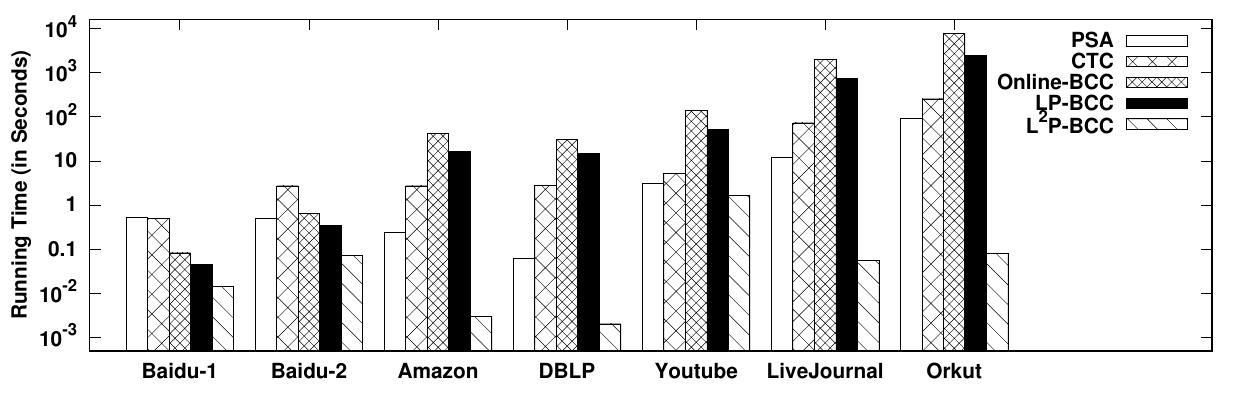}
	\vspace{-9mm}
	\caption{Efficiency evaluation on networks with ground-truth communities.}
	\vspace{-2mm}
	\label{fig.time}
\end{figure}

\spara{Queries and parameters.} First we generate a BCC query $Q=\{q_1, \ldots, q_\MultipleLabels\}$ and set $\MultipleLabels=2$ by default. For a default 2-labeled query $Q=\{q_l, q_r\}$, we set $k_1$ and $k_2$ as the coreness value of query vertices $q_l$ and $q_r$, and $b = 1$, respectively.
We vary two parameters, degree rank $Q_d$ and inter-distance $l$, to generate different query sets of pair vertices. 
For $Q_d$, we sort all vertices in ascending order of their degrees in a network. A vertex is regarded to be with degree rank of $X\%$, if it has top highest $X\%$ degree in the network. The default value of $Q_d$ is $80\%$, which means that a query vertex has a degree higher than the degree of $80\%$ vertices in the whole network. The inter-distance $l$ is the shortest path between two query vertices. The default $l = 1$ indicates that the two query vertices are directly connected in the network. 

\begin{figure*}[t!]
	\centering \mbox{
		\subfigure[\Quarter]{\includegraphics[width=0.2\linewidth]{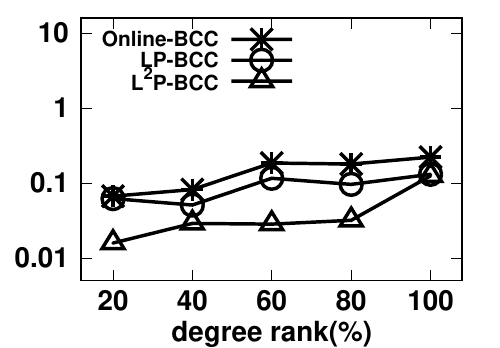}}
		\subfigure[\Year]{\includegraphics[width=0.2\linewidth]{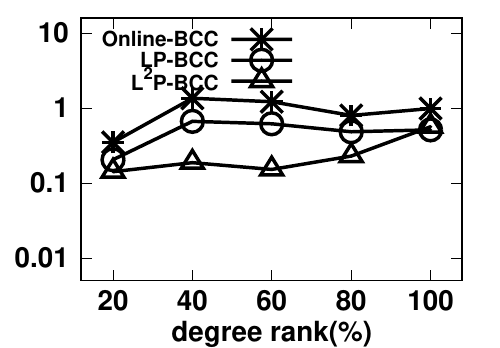}} 
		\subfigure[DBLP]{\includegraphics[width=0.2\linewidth]{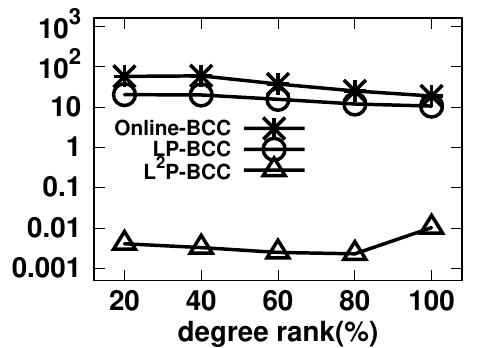}}
		\subfigure[LiveJournal]{\includegraphics[width=0.2\linewidth]{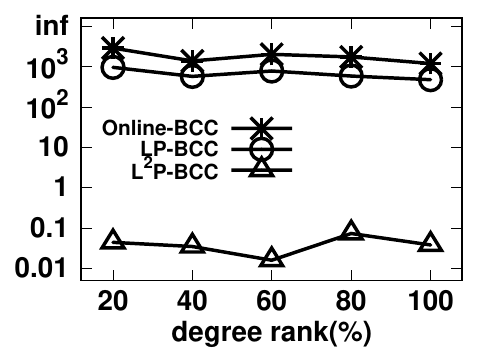}} 
		\subfigure[Orkut]{\includegraphics[width=0.2\linewidth]{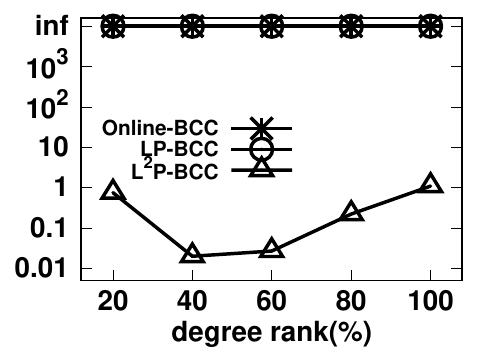}} 
		\vspace*{-0.5cm}
	}
	\vspace*{-0.5cm}
	\caption{Varying vertex degree rank: Query Time.}
	\vspace*{-0.4cm}
	\label{fig.d}
\end{figure*}
%
%
\begin{figure*}[t!]
	\centering \mbox{
		\subfigure[\Quarter]{\includegraphics[width=0.2\linewidth]{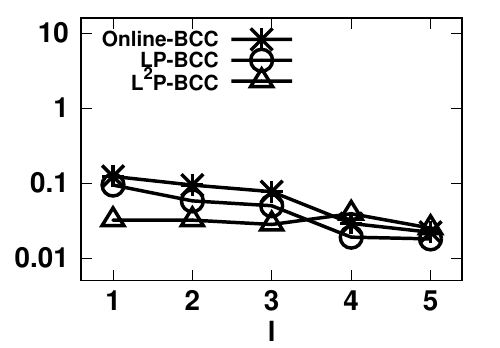}}
		\subfigure[\Year]{\includegraphics[width=0.2\linewidth]{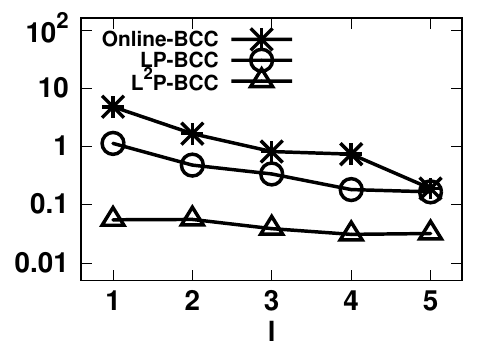}} 
		\subfigure[DBLP]{\includegraphics[width=0.2\linewidth]{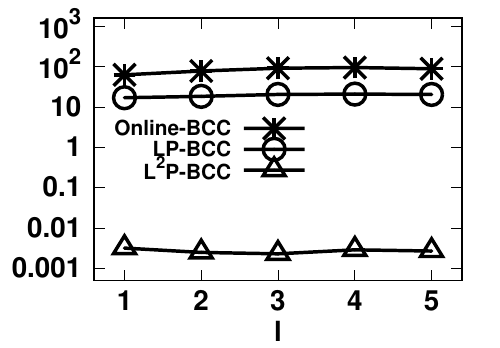}}
		\subfigure[LiveJournal]{\includegraphics[width=0.2\linewidth]{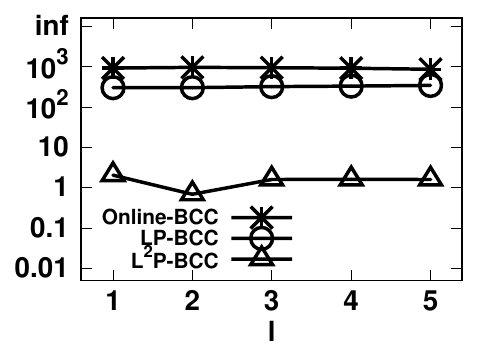}} 
		\subfigure[Orkut]{\includegraphics[width=0.2\linewidth]{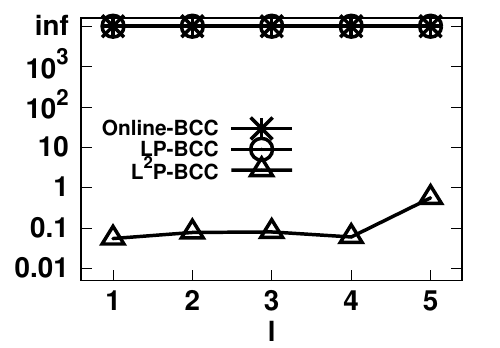}} 
		\vspace*{-0.5cm}
	}
	\vspace*{-0.5cm}
	\caption{Varying inner distance l: Query Time.}
	\vspace*{-0.4cm}
	\label{fig.l}
\end{figure*}

\begin{figure*}[t]
	\centering \mbox{
		\subfigure[\Quarter]{\includegraphics[width=0.2\linewidth]{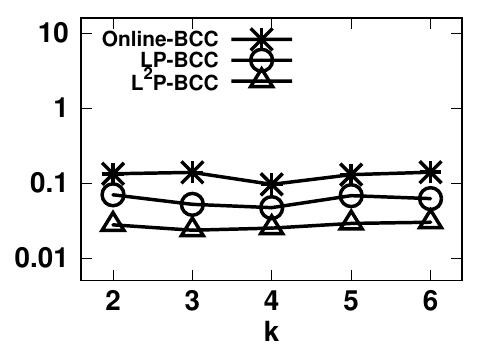}}
		\subfigure[\Year]{\includegraphics[width=0.2\linewidth]{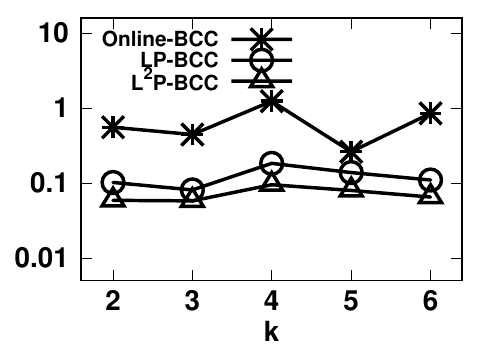}} 
		\subfigure[DBLP]{\includegraphics[width=0.2\linewidth]{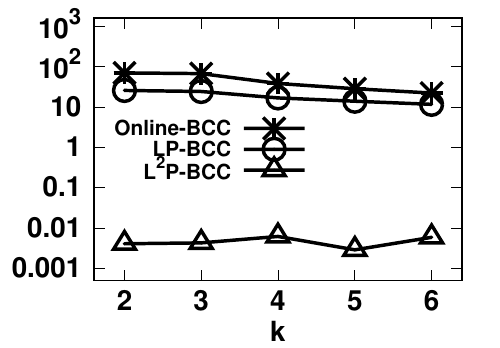}}
		\subfigure[LiveJournal]{\includegraphics[width=0.2\linewidth]{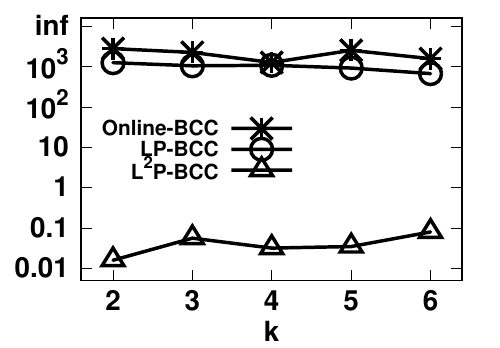}} 
		\subfigure[Orkut]{\includegraphics[width=0.2\linewidth]{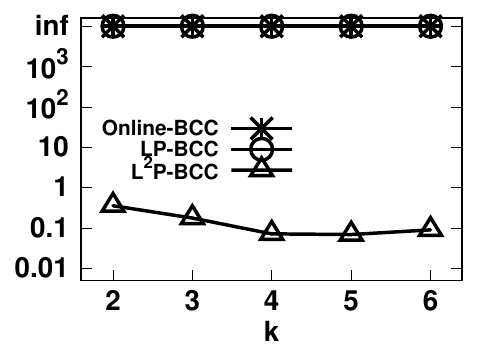}} 
		\vspace*{-0.5cm}
	}
	\vspace*{-0.5cm}
	\caption{Varying core value k: Query Time.}
	\vspace*{-0.4cm}
	\label{fig.k}
\end{figure*}
%
%
\begin{figure*}[t]
	\centering \mbox{
		\subfigure[\Quarter]{\includegraphics[width=0.2\linewidth]{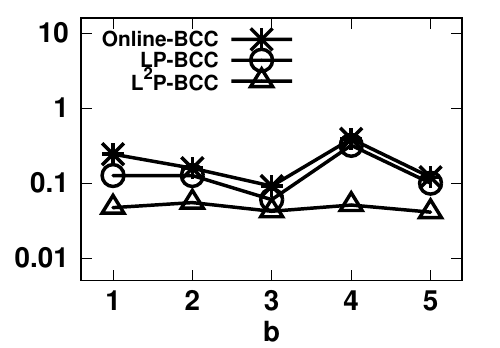}}
		\subfigure[\Year]{\includegraphics[width=0.2\linewidth]{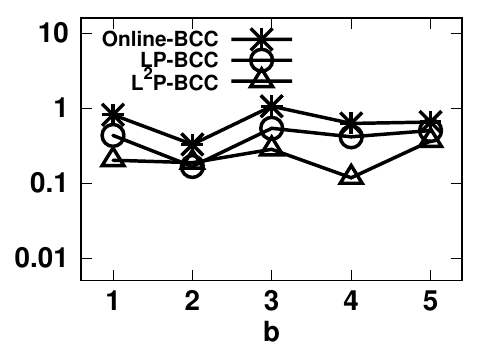}} 
		\subfigure[DBLP]{\includegraphics[width=0.2\linewidth]{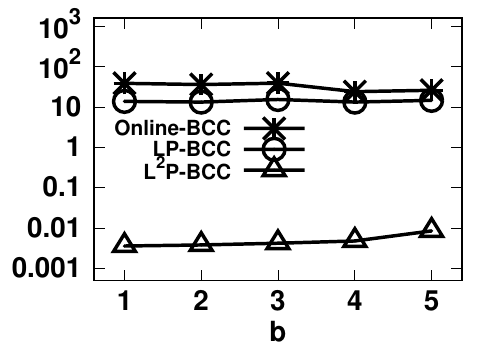}}
		\subfigure[LiveJournal]{\includegraphics[width=0.2\linewidth]{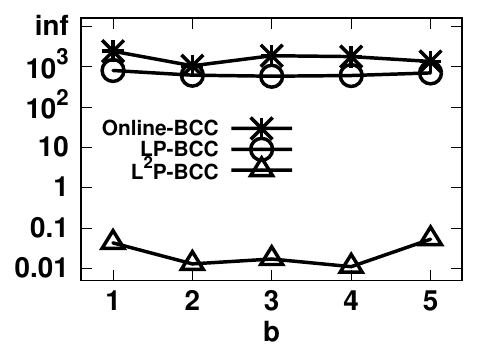}} 
		\subfigure[Orkut]{\includegraphics[width=0.2\linewidth]{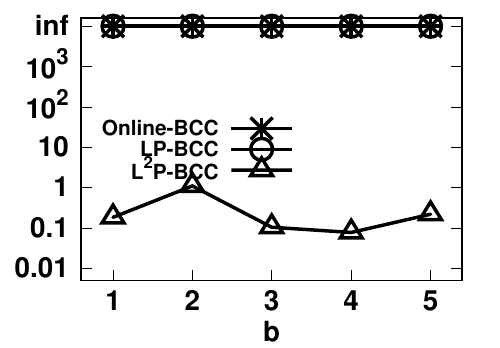}} 
		\vspace*{-0.5cm}
	}
	\vspace*{-0.5cm}
	\caption{Varying butterfly value b: Query Time.}
	\vspace*{-0.4cm}
	\label{fig.b}
\end{figure*}

\begin{figure*}[t]
	\centering \mbox{
		\subfigure[\Quarter]{\includegraphics[width=0.2\linewidth]{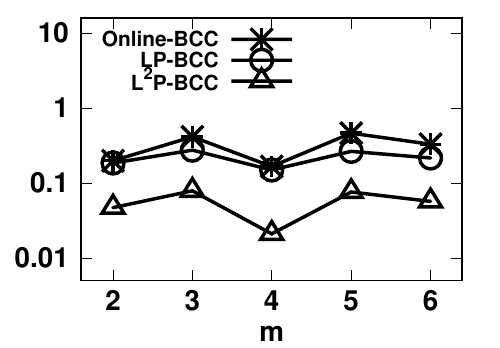}}
		\subfigure[\Year]{\includegraphics[width=0.2\linewidth]{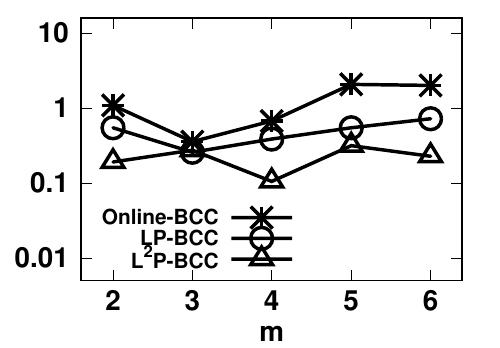}} 
		\subfigure[DBLP-M]{\includegraphics[width=0.2\linewidth]{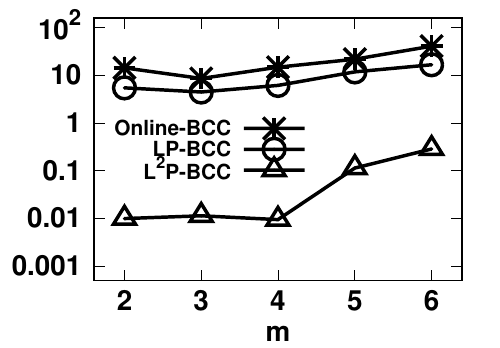}}
		\subfigure[LiveJournal-M]{\includegraphics[width=0.2\linewidth]{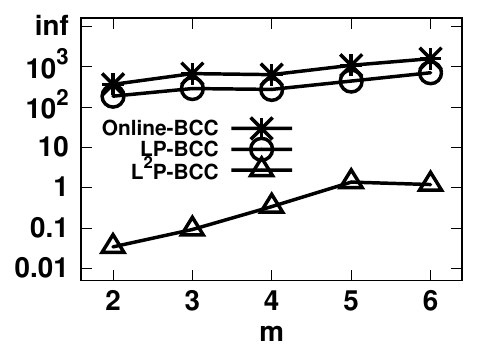}} 
		\subfigure[Orkut-M]{\includegraphics[width=0.2\linewidth]{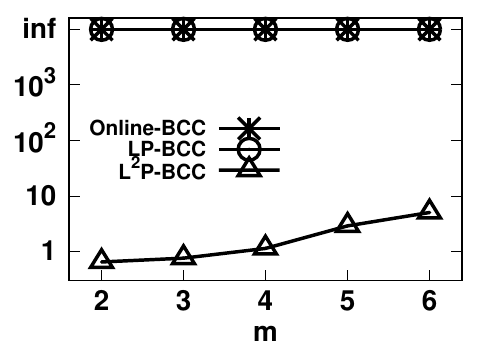}} 
		\vspace*{-0.5cm}
	}
	\vspace*{-0.5cm}
	\caption{Varying the number of query vertices with different labels $\MultipleLabels$: Query Time.}
	\vspace*{-0.4cm}
	\label{fig.multiple_efficiency}
\end{figure*}

\subsection{Quality, Efficiency, and Parameter Evaluations}

\spara{Exp-1: Quality evaluation with ground-truth communities.}
We evaluate the effectiveness of different community search models over labeled graphs. Figure~\ref{fig.f1} reports the averaged F1-scores of all methods over $1,000$ random queries on seven networks. We observe that our approaches achieve the highest F1-score on all networks against  \CTC \citep{huang2015approximate} and \PSA \citep{li2019efficient}. \LBCC is better than \BD and \BDQL on most datasets except for LiveJournal. All methods cannot have good results on Youtube.


\spara{Exp-2: Efficiency evaluation of all methods.}
We evaluate the efficiency performance of different community search models. Figure~\ref{fig.time} shows the running time results of all methods. All methods finish the queries processing within $30$ minutes, except \BD and \BDQL on Orkut as they generate a large candidate graph $G_0$. Most other methods use the local search strategy and run much faster than \BDQL. Interestingly, \BD and \BDQL run slightly faster than \CTC  and \PSA on \Quarter and \Year. Overall, \LBCC achieves the best efficiency performance, which can deal with one BCC search query within 1 second on most datasets.

\begin{table}[t]
	\centering
	\small
	\caption{A comparison of \BD and \BDQL on DBLP, in terms of query distance calculation and butterfly computation.
		The ``\#butterfly counting'' denotes the calling times of the butterfly counting procedure in Algorithm~\ref{algo:count}. Other steps are measured by their running time in seconds. }
	\vspace{-0.2cm}
	\label{ablation}
	\begin{tabular}{ l|rrrrr }
		\hline
		 Methods  & \BD & \BDQL & Speedup \\
		\hline
		Query distance calculation & $1.58$ & $0.75$ & 2.1\textbf{x}\\
		\hline
		Leader pair update & $4.98$ & $0.46$ & 10.8\textbf{x} \\
		\#butterfly counting & $34.86$ & $1.21$ & 28.8\textbf{x} \\ 
		\hline
		Total time  & $11.5$ & $4.04$ & 2.8\textbf{x} \\\hline
	\end{tabular}
	\vspace{-0.5cm}
\end{table}

\spara{Exp-3: Efficiency evaluation on different queries.}
To evaluate the efficiency of our proposed improved strategies, we compare \BD, \BDQL, and \LBCC. Figure~\ref{fig.d} and Figure~\ref{fig.l} show the running time on different networks, varying the vertex degree rank and inter-distance respectively. 
The most efficient method is \LBCC. With the increased vertex degree, \BD and \BDQL become faster on DBLP and LiveJournal in  Figure~\ref{fig.d}(c) and (d) as the vertices with a higher degree have denser and smaller induced graphs; it is different on \Quarter and \Year because of the dense structure of the graph. With the increased query vertices inter-distance $l$, the running time also increases on DBLP and LiveJournal. The reason may be that two query vertices are not the leader pair, leading to the low efficiency of our leader identification strategy to find the leaders satisfying the butterfly degree constraint.


\spara{Exp-4: Parameter sensitivity evaluation}. 
We evaluate the parameter sensitivity of $k_1, k_2, b$ on efficiency performance. 
When we test one parameter, the other two parameters are fixed. Moreover, we test one parameter of $k_1$ and $k_2$ as $k$ due to their symmetry parameter property. 
Figure~\ref{fig.k} shows the running time by varying the core value $k$. We observe that the larger $k$ results in less running time since it generates a smaller $G_0$ for a larger $k$. 
Figure~\ref{fig.b} shows the running time by varying the butterfly degree $b$. We observe that our approach achieves a stable efficiency performance on different values $b$.

\spara{Exp-5: Efficiency evaluations of query distance and butterfly computations}. 
We evaluate the efficiency of our proposed fast strategies in query distance computation in Algorithm~\ref{algo:sssp} and leader pair identification in Algorithms~\ref{algo:LeaderPair} and~\ref{algo:UpdatePair}. We compare \BDQL and \BD on DBLP using 1000 queries and report the detailed results in Table~\ref{ablation}. Algorithm~\ref{algo:sssp} achieves $2.1$\textbf{x} speedups than the baseline of query distance computation by \BDQL. 
The leader pair identification in Algorithms~\ref{algo:LeaderPair} and~\ref{algo:UpdatePair} greatly reduces the calling times of butterfly counting, validating that our proposed leader pair identifications can target a pair of stable leaders with large butterfly degrees even for graph removals. The leader pair identification achieves $28.8$\textbf{x} and $10.8$\textbf{x} speedups on \#butterfly counting and the running time of leader pair update, respectively. In terms of total running time, \BD integrating fast strategies achieves $2.8$\textbf{x} speedups against \BDQL.

\subsection{Three Real-world Case Study Comparisons}\label{sec:case_study_sec}
To evaluate the effectiveness of the BCC model, we conduct case studies on three real-world networks including a global flight network~\footnote{\scriptsize{https://raw.githubusercontent.com/jpatokal/openflights/master/data/routes.dat}}, an international trade network~\footnote{\scriptsize{https://wits.worldbank.org/datadownload.aspx?lang=en}} and the J. K. Rowling's Harry Potter series fiction network~\footnote{\scriptsize{https://github.com/efekarakus/potter-network}}. We compare our method \BDQL with the closest truss community search (\CTC \citep{huang2015approximate}). We set the parameter $b=3$ and the values of $k_1, k_2$ as the coreness of query vertices $q_l$ and $q_r$ for \BDQL. The method \CTC uses its default setting~\citep{huang2015approximate}. 
%

\begin{figure}[t]
	\centering \mbox{
		\subfigure[Our Butterfly-Core Community]{\includegraphics[width=0.65\linewidth, height=2.5cm]{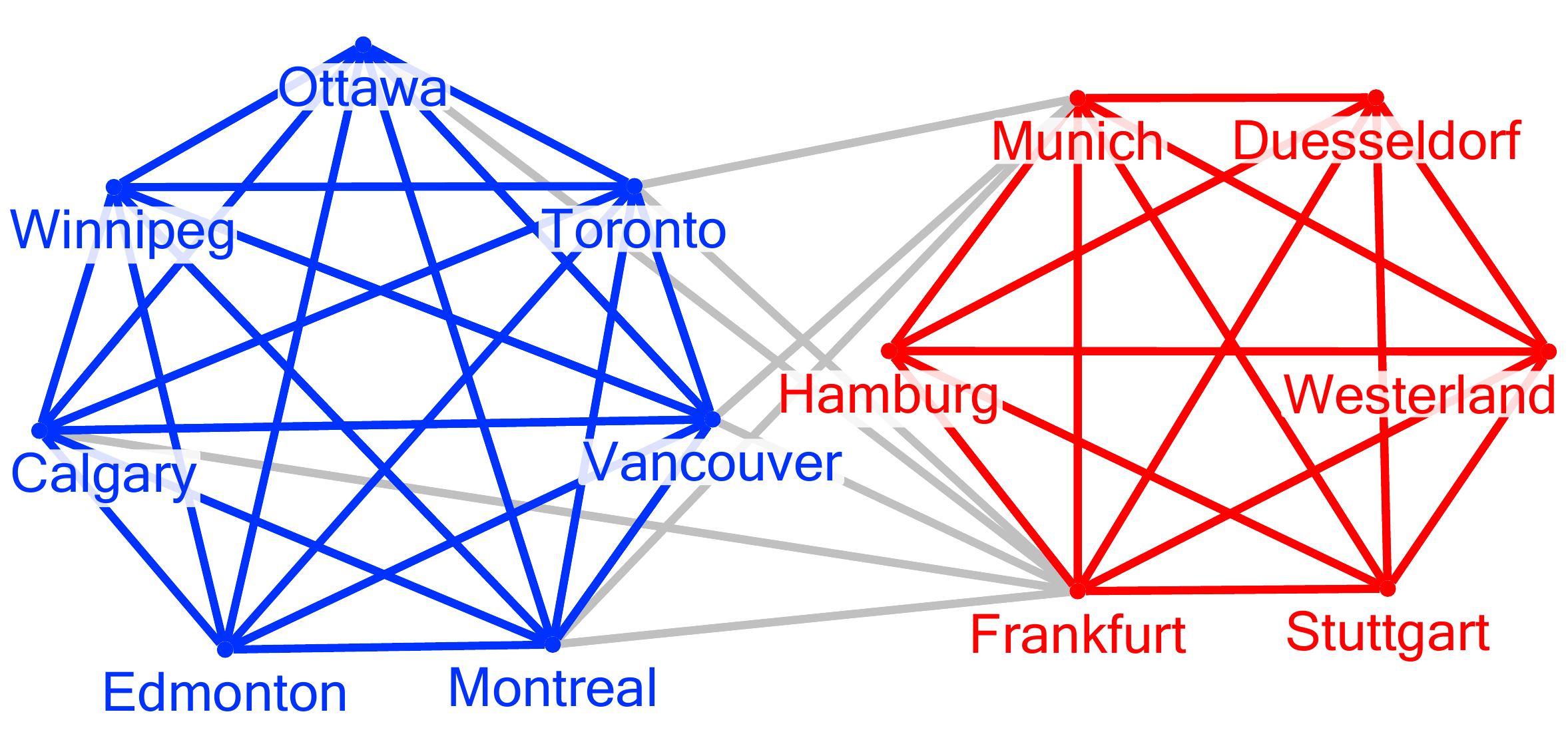}}
		\subfigure[CTC~\citep{huang2015approximate}]{\includegraphics[width=0.35\linewidth, height=2.5cm]{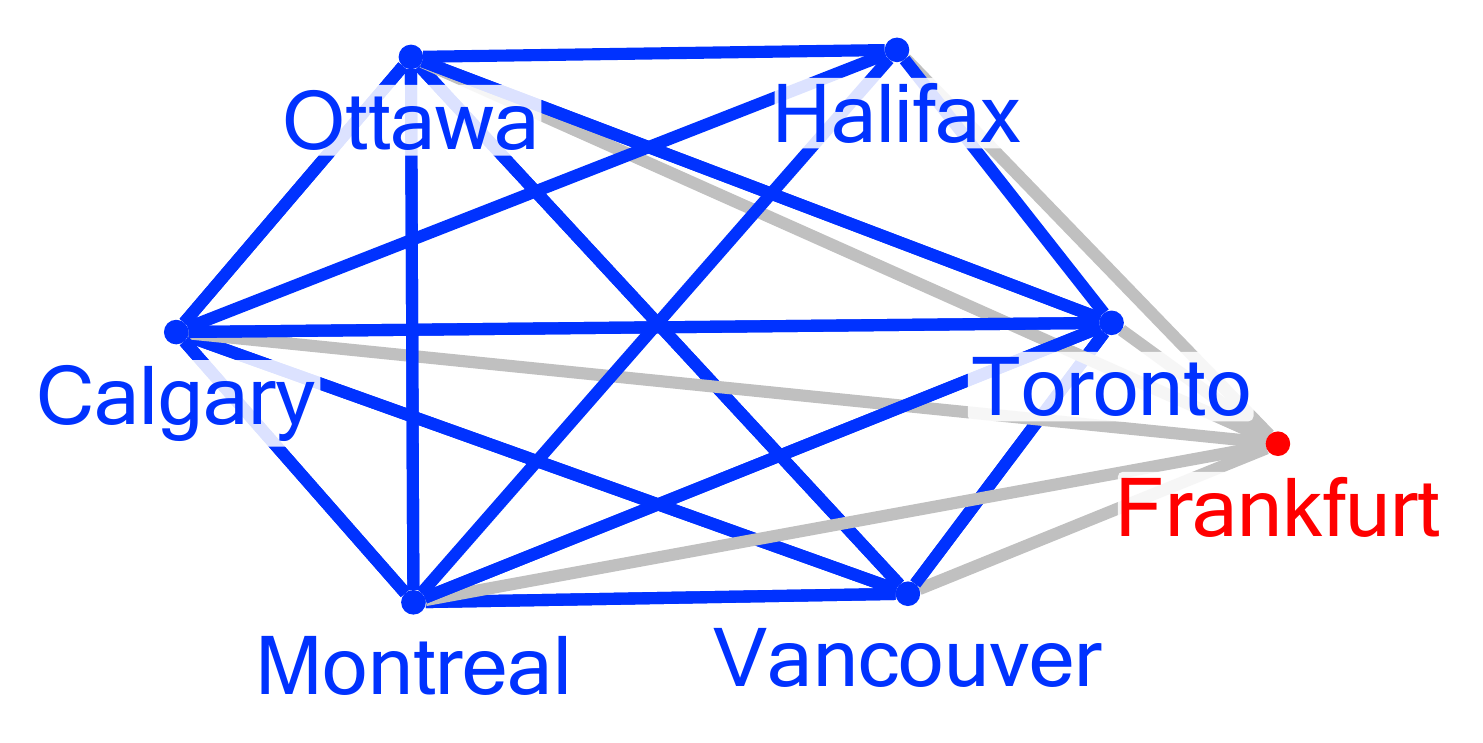}} 
		\vspace{-0.5cm}
	}
	\vspace{-0.5cm}
	\caption{Case study of two community search models on flight networks for query $Q=$\{``Toronto'', ``Frankfurt''\}.}
	\vspace{-0.5cm}
	\label{fig.flight}
\end{figure}

\stitle{Exp-6: A case study on flight networks}. Here we study a labeled graph of flight network with $238$ different labels, where each vertex represents a city with a label of the home country. The flight network has $3,334$ vertices and $19,205$ edges. Note that this graph is an undirected single-edge graph generated from the source data$^{1}$, which generates an edge between two cities if there exists more than one airline between them. 
Thus, edges between the same labeled vertices are domestic airlines. On the other hand, international airlines are represented by the edges between different labeled vertices. 
We set the query vertices $Q=$\{``Toronto'', ``Frankfurt''\} to discover the dense flight transportation networks between them. The cities ``Toronto'' and ``Frankfurt'' have a label of ``Canada'' and ``Germany'' respectively. Figure~\ref{fig.flight}(a) depicts our result of BCC. The flight community consists of a $6$-core in blue, butterfly degree is $3$ in gray, and a $5$-core in red. A butterfly forms by four cities ``Toronto'', ``Vancouver'', ``Frankfurt'', and ``Munich'', which are the transport hubs for transnational airlines. The core subgraphs in blue and red reflect the complex and dense networks of domestic airlines in Canada and Germany.
Figure~\ref{fig.flight}(b) shows the community result of \CTC~\citep{huang2015approximate}. 
It consists of a dense airline community involving ``Toronto'' and ``Frankfurt''. However, most discovered vertices in Figure~\ref{fig.flight}(b) are the cities in Canada, which fails to find the international airline community as our BCC model in Figure~\ref{fig.flight}(a). 

\begin{figure}[t]
	\centering \mbox{
		\subfigure[Our Butterfly-Core Community]{\includegraphics[width=0.65\linewidth, height=2.5cm]{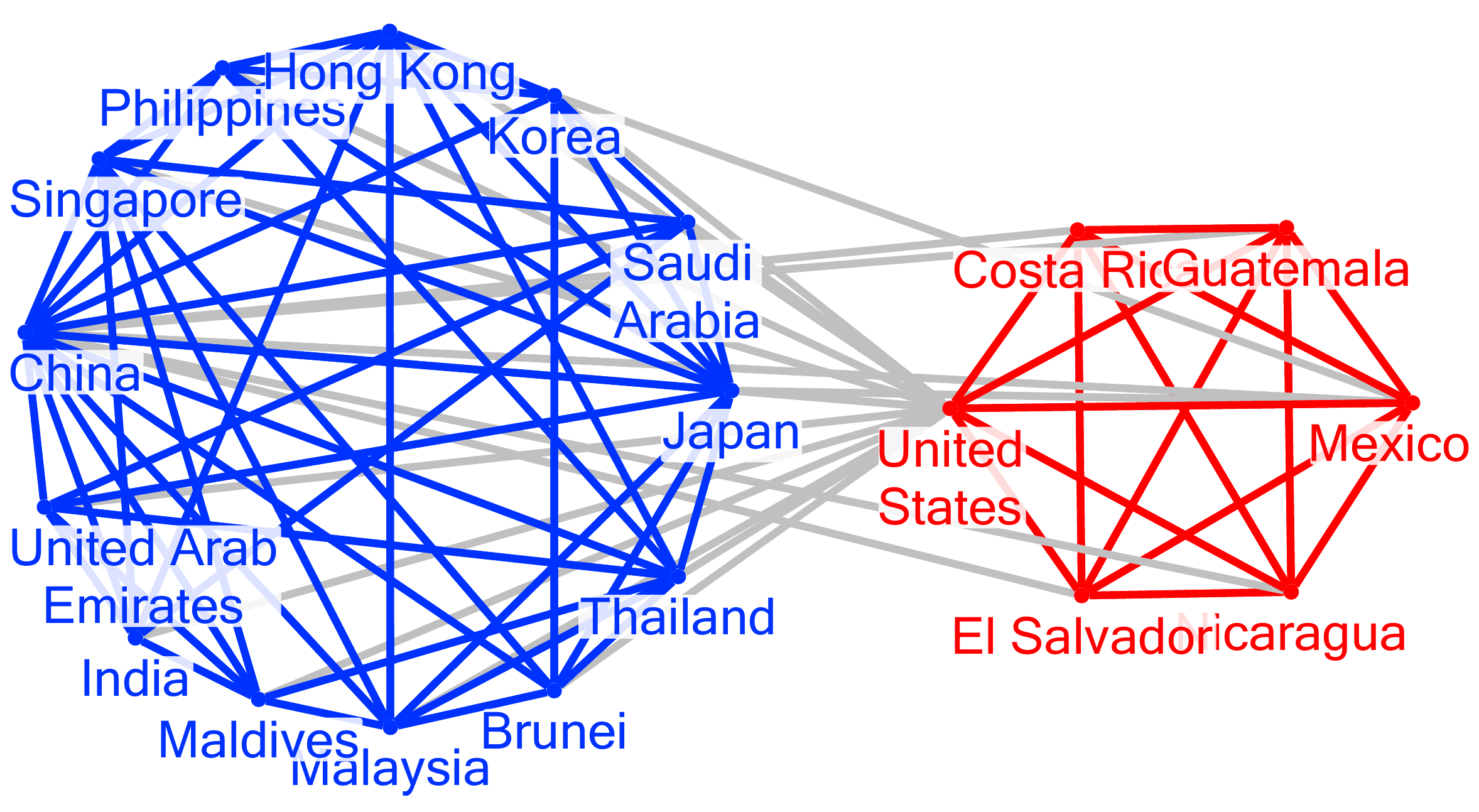}}
		\subfigure[CTC~\citep{huang2015approximate}]{\includegraphics[width=0.35\linewidth, height=2.5cm]{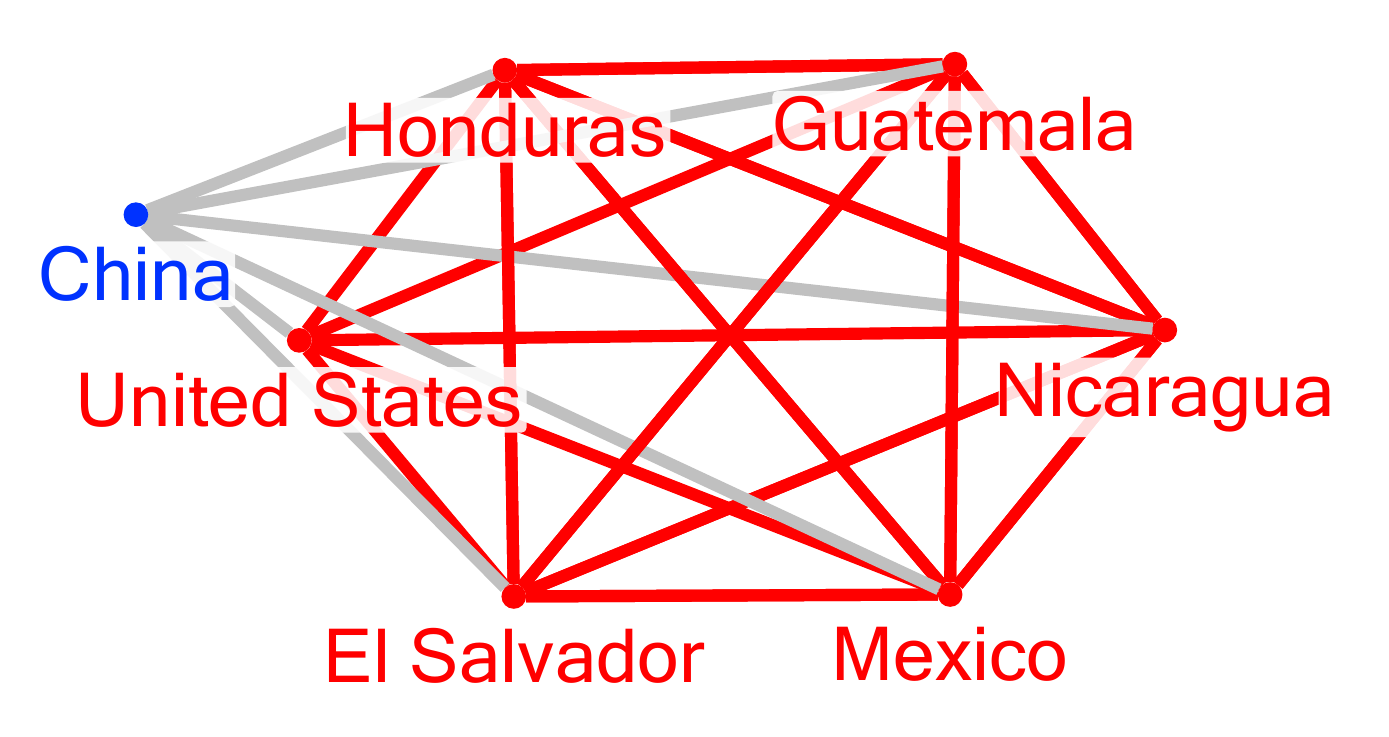}}
	}
	\vspace{-0.5cm}
	
	\caption{Case study of two community search models on trade networks for query $Q=$\{``United State'', ``China''\}.}
	\vspace{-0.5cm}
	\label{fig.trade}
\end{figure}

\begin{figure}[t]
	\centering \mbox{
		\subfigure[Our Butterfly-Core Community]{\includegraphics[width=0.65\linewidth, height=2.5cm]{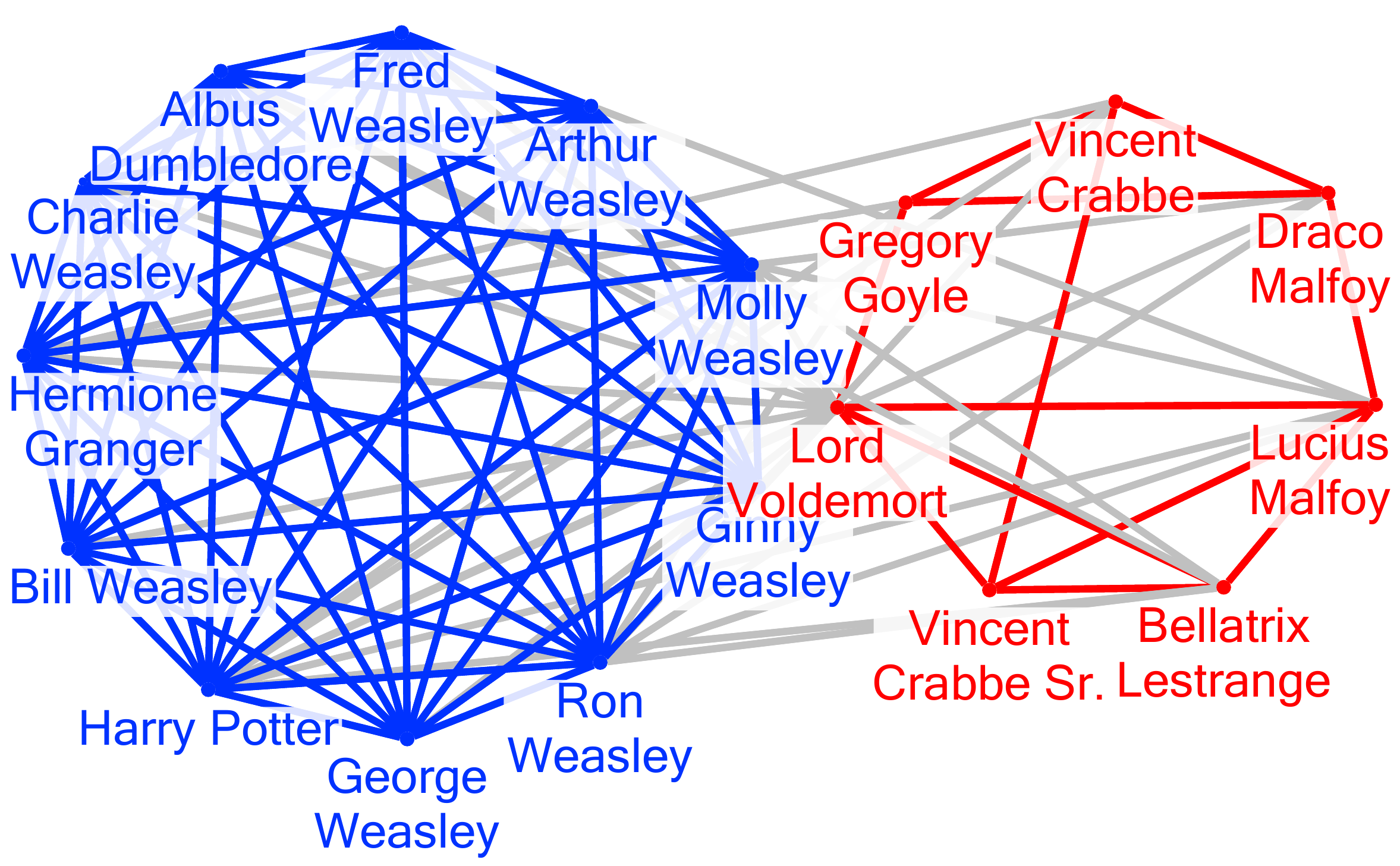}}
		\subfigure[CTC~\citep{huang2015approximate}]{\includegraphics[width=0.35\linewidth, height=2.5cm]{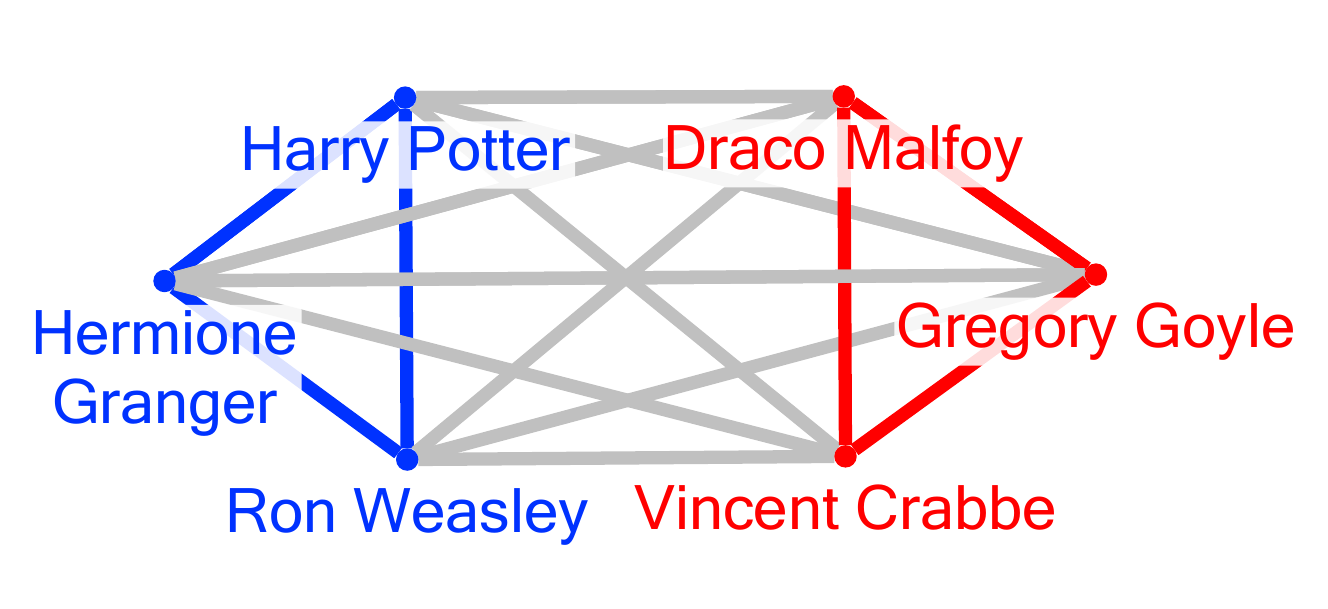}} 
	}
	\vspace{-0.5cm}
	\caption{Case study of two community search models on Harry Potter fiction networks for query $Q=$\{``Ron Weasley'', ``Draco Malfoy''\}.}
	\vspace{-0.6cm}
	\label{fig.harry}
\end{figure}

\stitle{Exp-7: A case study on trade networks}. The international trade network describes trade relations between countries/regions. Each vertex represents a country/region with its located continent as a label, e.g., the label of ``China''  is ``Asia''. The trade labeled graph has seven labels. We add an edge between two countries/regions if one is the top-$5$ import or export trade partners to the other in $2019$. There are $249$ vertices and $1,281$ edges. We set query vertices $Q = $\{``United State'', ``China''\}. Figure \ref{fig.trade}(a) reports our BCC community result. It consists of dense trade subnetworks in ``Asia'' and ``North America'', and also the trade leaders, i.e., ``United State'' and ``China'' have the most transcontinental trades. Figure \ref{fig.trade}(b) shows the \CTC community~\citep{huang2015approximate} fails to find the other major trade partners in ``Asia''. 



\stitle{Exp-8: A case study on Harry Potter fiction networks}. The Harry Potter network is a $2$-labeled graph, where each vertex represents a character. Each character has a label representing his camp justice or evil. An edge is added between two characters if they have intersections in the fiction. The edges between two same labeled vertices denote the family and ally in the same camp, while the edges between two different labeled vertices denote hostility enemies in different camps. There are $65$ vertices and $513$ edges. We set the query $Q=$\{``Ron Weasley'', ``Draco Malfoy''\}. Figure \ref{fig.harry}(b) shows that the \CTC community only finds  ``Ron Weasley'' with his best friends of Harry and Hermione in the Gryffindor house, Malfoy and his two cronies belonging in the Slytherin house. However, it misses the main leader in the evil camp, i.e., ``Lord Voldemort''. What's more, it also fails to find Ron's huge family, including his parents and brothers, as our BCC result is shown in Figure~\ref{fig.harry}(a).

In summary, existing community models~\citep{huang2015approximate, li2019efficient} which ignore the graph labels, are hard to discover those communities across over two different labeled groups, which make a huge difference from our BCC model in three aspects: (1) they ignore the different dense level of two labeled groups; (2) they confuse the semantics of different edge types (e.g., international and domestic airlines); (3) they may miss some members of one group due to the strong structural required by models. On other hand, our BCC model aims to mine out the whole dense teams including the leader pairs, inner edges within a group, and cross edges between two groups. 

\begin{figure}[t]
	\centering \mbox{
		\subfigure[\Quarter]{\includegraphics[width=0.5\linewidth, height=2.5cm]{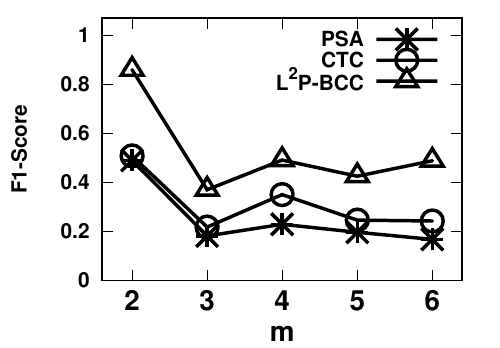}}
		\subfigure[\Year]{\includegraphics[width=0.5\linewidth, height=2.5cm]{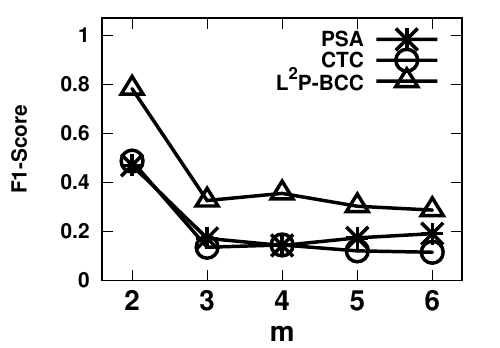}} 
	}
	\vspace{-0.5cm}
	\caption{Quality evaluation on \Quarter and \Year with multi-labeled ground-truth communities.}
	\vspace{-0.6cm}
	\label{fig.f1_multiple}
\end{figure}


\begin{figure*}[t]
	\centering \mbox{
		\subfigure[2-labeled BCC]{\includegraphics[width=0.4\linewidth, height=4cm]{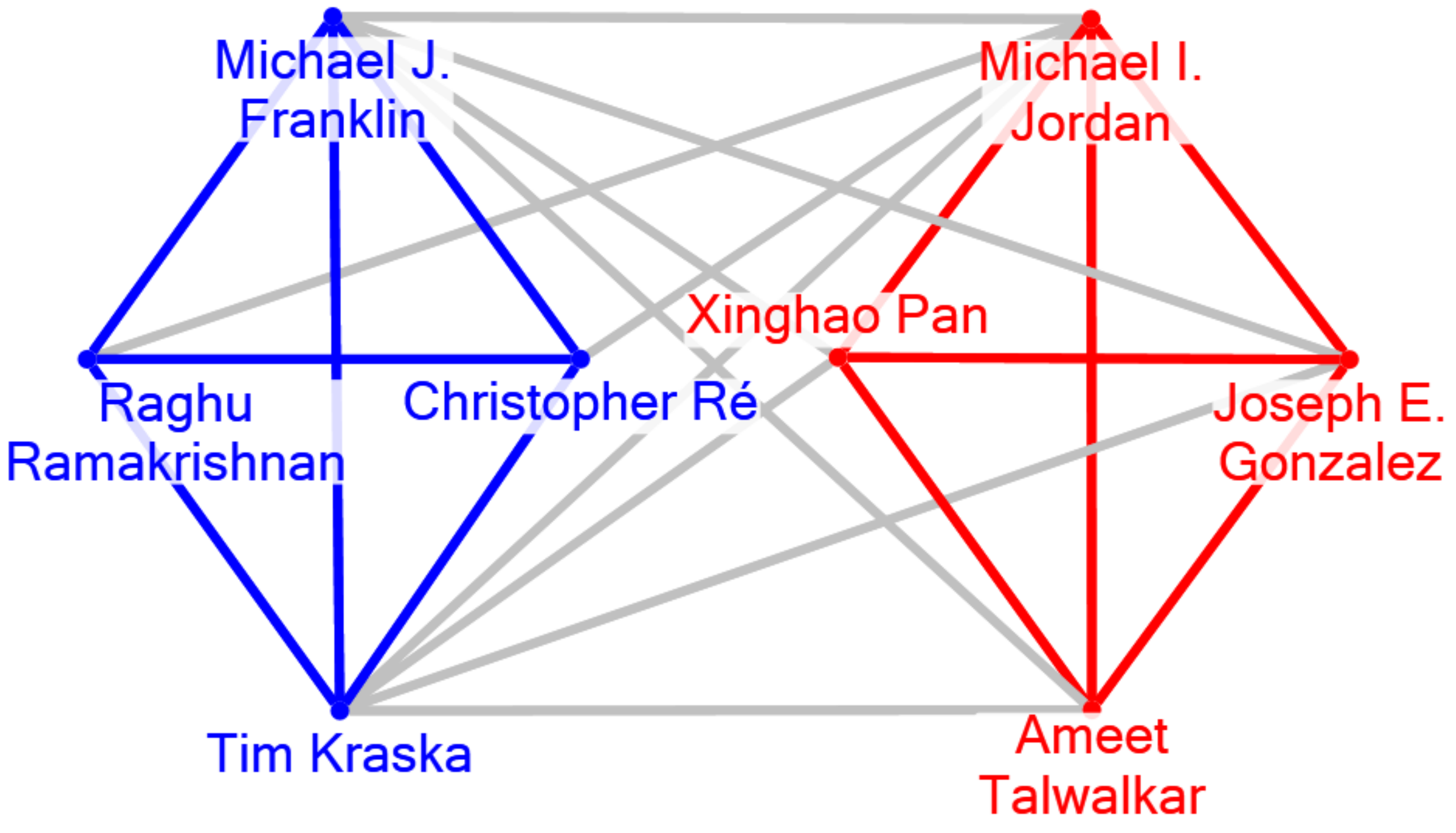}}  
		\subfigure[3-labeled BCC]{\includegraphics[width=0.6\linewidth, height=4cm]{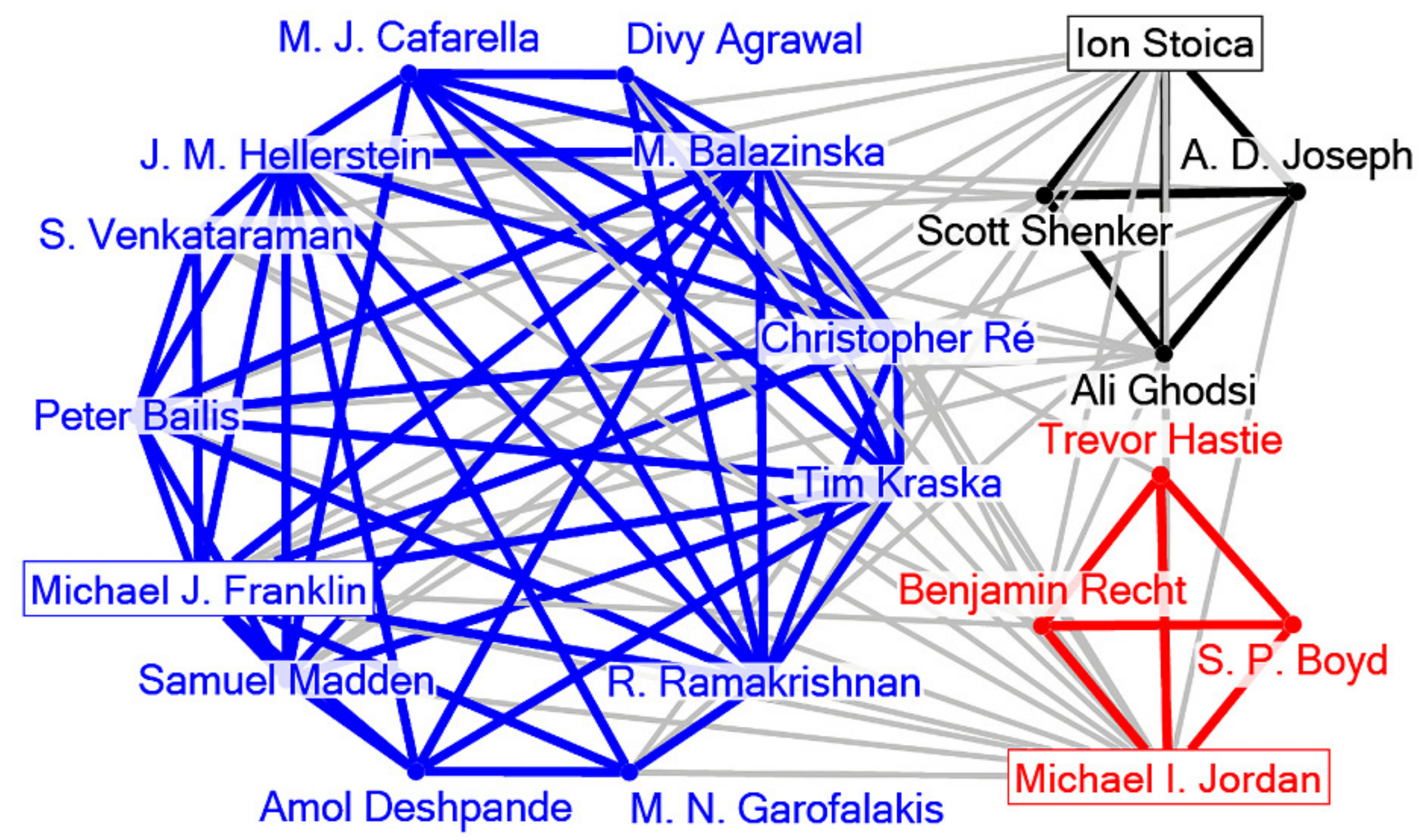}} 
	}
	\vspace{-0.5cm}
	\caption{Case study of multi-labeled BCC search on academic collaboration networks for interdisciplinary research group discovery using two queries $Q_1$ and $Q_2$. Here, $Q_1 =$ \{``Tim Kraska'', ``Michael I. Jordan''\} and $Q_2 =$ \{``Michael J. Franklin'', ``Michael I. Jordan'', ``Ion Stoica''\}.}
	\label{fig.multiple_case_studies}
\end{figure*}

\subsection{Multi-labeled BCC Search Evaluation} 
\spara{Exp-9: Quality evaluation of multi-labeled BCC search with ground-truth communities.}
We conduct quality evaluations of multi-labeled BCC search on two datasets of \Quarter and \Year with multi-labeled ground-truth communities, representing joint projects among multiple department teams in the enterprise.
We vary the number of query labels $\MultipleLabels$ in $Q$ and report F1-score on the average of 100 queries. Figure~\ref{fig.f1_multiple} reports the results of  \CTC \citep{huang2015approximate}, \PSA \citep{li2019efficient}, and 
our local method \LBCC for multi-labeled BCC search in Algorithm~\ref{algo:Multiple}. We observe that all methods achieve worse performance with the increased query labels $\MultipleLabels$, indicating that a more challenging task for identifying multi-labeled BCC communities. Nevertheless, our approach 
\LBCC consistently performs better than the competitors \CTC and \PSA for all parameters $\MultipleLabels$ on Figures 12(a) and 12(b).


\spara{Exp-10: Efficiency evaluation of multi-labeled BCC search.}
We evaluate the efficiency of multi-labeled BCC search methods. Following the mBCC search framework in Algorithm~\ref{algo:Multiple}, we implement three extension approaches \BD, \BDQL and \LBCC for multi-labeled BCC search accordingly. 
Beside two ground-truth datasets \Quarter and \Year, we further generate and use three large graphs with multiple labels. Specifically, we assign six vertex labels for all vertices randomly in graphs DBLP, LiveJournal and Orkut, denoted as DBLP-M, LiveJournal-M, and Orkut-M respectively.  
Figure~\ref{fig.multiple_efficiency} shows the running time of 
three mBCC extension methods \BD, \BDQL and \LBCC, varying by the number of query labels. Figures~\ref{fig.multiple_efficiency}(a)-(b) show that all methods achieve stable efficiency performance for  different query labels  $\MultipleLabels$ on small graphs \Quarter and \Year. On the other hand, all methods takes a longer running time slightly with the increased $\MultipleLabels$ on large graphs in Figures~\ref{fig.multiple_efficiency}(c)-(d). The reason is that mBCC search takes more cost of shortest path computation for $\MultipleLabels$ queries, confirming the time complexity analysis of Algorithm~\ref{algo:Multiple} in Section~\ref{sec:multiple}. Importantly, our local method \LBCC once again runs fastest among all the three methods, validating the effectiveness of our fast strategies even for multi-labeled BCC search.



\spara{Exp-11: Case study of multi-labeled BCC search for interdisciplinary collaboration groups on DBLP.} 
We construct a real-world dataset of research collaboration network based on the ``DBLP-Citation-network V12'' at Aminer~\citep{tang2008arnetminer}\footnote{\scriptsize{https://www.aminer.cn/citation}}. The collaboration network has $144,334$ vertices, $1,821,930$ edges, and a total of $7$ vertex labels, which is publicly available\footnote{\scriptsize{https://github.com/zhengdongzd/butterfly-core}}. Each vertex represents an author. Two authors have an edge if they have at least one paper collaboration. For each vertex, we count on the author's published papers based on research topics, e.g., ``Database'' papers in the venues of SIGMOD, VLDB, and so on. Finally, we take the \emph{vertex label} of an author as the research field of his/her most published papers, e.g., ``Database'', ``Machine Learning'', and so on. An edge between two authors with different research field labels, is regarded as an \emph{ interdisciplinary collaboration}. We set the parameter $b = 3$ and the coreness values of $k_i=3$ for all query vertices $q_i$.
First, we query the 2-labeled BCC community containing $Q_1 = $\{``Tim Kraska'', ``Michael I. Jordan''\}, which is shown in Figure~\ref{fig.f1_multiple}(a). This is an interdisciplinary research group crossover the fields of ``Database'' and ``Machine Learning'', which works on the machine learning techniques for database systems and vice versa (a.k.a. ML4DB and DB4ML). Homogeneous groups in red and blue are densely connected, forming the 3-core and indeed 4-clique, respectively. In terms of butterfly structure in bipartite graphs, two query vertices ``Tim'' and ``Michael'' have a butterfly degree of $6$ and $3$ respectively, which has interdisciplinary collaborations with other research group and bridges these two research communities together. 
Second, we conduct multi-labeled BCC search, by querying three labeled vertices $Q_2 = $\{``Michael J. Franklin'', ``Michael I. Jordan'', ``Ion Stoica''\}, who are from the research fields of ``Database'', ``Machine Learning'', and ``Systems and Networking'', respectively. The result of $3$-labeled cross-discipline community is shown in Figure~\ref{fig.f1_multiple}(b). As can be seen, several well-known scholars of these fields appear in the BCC community and have close intra-group collaborations and interdisciplinary cooperations across different research fields of ``Machine Learning'', ``Database'', and ``Systems and Networking''. The database group is a $3$-core and there are $13$ vertices, the cross-group connectivity is formed by the \emph{cross-group path} $P=$\{``Machine learning'', ``Database'', ``Systems and Networking''\}, since  the database group both has cross-group interactions with ``Machine Learning'' and ``Systems and Networking'' groups.

\section{Conclusion}\label{sec:con}

In this paper, we studied a new problem of butterfly-core community search over a labeled graph. We theoretically proved the problem's NP-hardness and non-approximability. To tackle it efficiently, we proposed a $2$-approximation solution for BCC search. To further improve the efficiency, we developed a fast local method L$^2$P-BCC to quickly calculate query distances and identify leader pair. We proposed the mBCC model and the extended framework to explore cross-group community search with multiple labels. Extensive experiments on seven real-world networks with ground-truth communities and interesting case studies validated the effectiveness and efficiency of our BCC and mBCC models and algorithms.


\bibliographystyle{ACM-Reference-Format}
\bibliography{butterfly}

\end{document}